\documentclass[titlepage,aps,pre,twocolumn,amsmath,amssymb,superscriptaddress,reprint,longbibliography,nofootinbib]{revtex4-1}

\usepackage[utf8]{inputenc}
\usepackage{graphicx}
\usepackage{color}
\usepackage[colorlinks=true,citecolor=blue]{hyperref}
\usepackage{units}
\usepackage{amsmath}
\usepackage{tabularx}
\usepackage{lipsum}
\usepackage{float}

\usepackage{amssymb}
\usepackage{xcolor}
\usepackage{extarrows}
\usepackage{tikz-cd}
\usepackage{calrsfs}

\DeclareMathOperator{\sign}{sign}

\newcommand{\bSigma}{\Sigma}
\newcommand{\bom}{\boldsymbol{m}}
\newcommand{\rect}{\text{rect}}

\AtBeginDocument{%
 \newwrite\bibnotes
 \def\bibnotesext{Notes.bib}
 \immediate\openout\bibnotes=\jobname\bibnotesext
 \immediate\write\bibnotes{@CONTROL{REVTEX41Control}}
 \immediate\write\bibnotes{@CONTROL{%
 apsrev41Control,author="08",editor="1",pages="1",title="0",year="1"}}
 \if@filesw
 \immediate\write\@auxout{\string\citation{apsrev41Control}}%
 \fi
}%

\begin{document}

\title{Geometrically frustrated systems which are as singles hotter than in company }
\author{Wolfgang Rudolf Bauer}
\email{bauer\_w@ukw.de}
\affiliation{
University/University Hospital of W\"urzburg (ZIM),
Oberd\"urrbacher Stra{\ss}e 6,
D-97080 W\"urzburg, Germany, Tel. +49-931-201-39011
}
\altaffiliation[Also at: ]{Comprehensive Heart Failure Center, Am Schwarzenberg 15, A15, D-97080 W\"urzburg, Germany} 

\date{\today}

\begin{abstract}
We show that a set of thermally weakly coupled geometrically frustrated systems (GFSs), each of which is  constraint to reside at negative Boltzmann temperatures, is in equilibrium cooler than its  constituents. It may even exhibit positive temperatures at low energies. The challenge for the second law of thermodynamics arising from potential heat flow related to the gradient of temperatures between a GFS and its environment is resolved by considering the energy fluctuations above the ground state. They are comprised in the canonical temperature, derived from information theory. Whereas the gradient of Boltzmann temperatures gives the direction of the stochastic drift of the most probable state of a GFS within its environment, the
canonical temperature gradient defines that of heat flow.

\begin{description}
\item[PACS numbers] 5.70.Ln, 5.20.Gg
\end{description}            
\end{abstract}
\maketitle

\section{Introduction}
Our  everyday experience is that systems of same temperature, which are brought from initial isolation to  thermal coupling, maintain their temperature, which is equivalent to that of the combined system. Absence of temperature gradients implies no heat flow, and the combined system is said to be in thermal equilibrium. Any other scenario would be in conflict with our notion of thermal stability and the second law of thermodynamics in Clausius's version \cite{clausius1872vi}\footnote{Exceptions of this thermal stability exist for intensive parameters as magnetization, where subsystems,  after brought in weak thermal contact, evolve towards different phases \cite{ramirez2008systems}.}. This base for thermometry derives  in phenomenological thermodynamics from the additive behavior of the state variables energy and  entropy, which  makes temperature of the combined system an intensive parameter, i.e. it is  equivalent to that of its subsystems. Entropy of the combined system has in this state its maximum \cite{callen1985thermodynamics}, which ensures a stable thermodynamic equilibrium.  
In statistical mechanics Boltzmann's concept of entropy predicts this state as the most probable one, which in ``normal'' macroscopic  systems \cite{Kubo} clearly dominates over any other state, and ensures that entropy becomes additive and temperature intensive. This concept may be consistently extended to systems capable of having negative temperatures \cite{baldovin2021statistical}. 

We recently described generic geometrically frustrated spin systems (GFSs), in which the ground state exhibits maximum degeneracy and Boltzmann entropy almost declines linearly with energy which makes the corresponding Boltzmann spin temperature constantly negative throughout \cite{bauer2022geometrically}. When brought in thermal contact with another system, the framework of Boltzmann entropy and temperature with reference to thermodynamic equilibrium of the combined system, i.e. principle of equal weight of microstates and detailed balance, may be straightforwardly transferred. The temperature gradient between a GFS and another system acts as a thermodynamic driving force which directs the combined system towards a more probables state. Boltzmann entropy of a single GFS declines almost linearly with energy, which implies a constant negative temperature. This linear dependence of entropy on energy implies that within a set of thermally weakly coupled GFSs, all combinations of energy, under the constraint of energy conservation, become equally probable. This indifferent rather than the stable equilibrium in``normal'' systems, implies that entropy of the assembly of GFSs is strongly superadditive, and, consequently, its temperature should differ from that of its constituents. At a first glance, this scenario would be in  conflict with the second law of thermodynamics, which relates heat flow to temperature gradients. In this article we systematically address how temperature of a set of thermally weakly coupled GFSs is related to that of its constituents, and how the notion that temperature gradients drive heat flow must be specified, to maintain the second law of thermodynamics.

\section{Weakly coupled GFSs as microcanonical ensemble}
\subsection{Single GFS}
The GFS we have in focus is a special Ising model of $N$ spins, with equal mutual anti-ferromagnetic interaction of strength $J$.  
The paradigm is that of primitive geometrically frustrated systems forming spin-ice \cite{moessner2001,moessner2006geometrical}. Within the GFS, the spins can be thought to be located at the $N$ vertices of a regular $N-1$-dimensional simplex. With orientation $s_i=\pm 1 $ of the $i$-th spin, the microstate of a GFS becomes  $\sigma=(s_1,\hdots,s_N)^T $, and the Hamiltonian reads
\begin{eqnarray}
H(\sigma)&=&\frac{J}{2}\sum_{i,j\;i\neq j}^{N} s_i s_j=\frac{J}{2}\; \left(m^2-N\right)\;, \text{with}\label{Eq:H0}\\
m&=&\sum_{i=1}^{N} s_i\;.\label{Eq:magnetization}
\end{eqnarray} 
as magnetic dipole moment. The offset $- J/2\; N $ is omitted furtheron. 
The set of microstates, $\bSigma_m$, with  magnetic moment $m $ has    
\begin{equation}
|\bSigma_m|=\binom{N}{\tfrac{N+m}{2}}\approx\tfrac{1}{\sqrt{2\pi}}\;\frac{2^N}{\sqrt{N/4}}\;e^{-\tfrac{1}{2}\;\left(\nicefrac{m^2}{N}\right)}\;\label{Eq:Energy}\;
\end{equation} 
% exploit from Gaussian approx of Binomial:
%\binom{N}{\tfrac{N+m}{2}}\approx \tfrac{1}{\sqrt{2\pi}}
% 1/\sqrt{N 1/2 1/2} \exp(-[(N+m)/2-N/2]^2/(2N 1/2 1/2) ) 2^N 
elements, where we assumed $N$ to be so large, that the Gaussian approximation of the binomials holds, and the dipole moments $m$ lies within several standard deviations $k\times\sqrt{N}$ so that $k/\sqrt{N}\ll1 $. With energy $E(m)=J/2 \; m^2 $, the  Boltzmann entropy $S_{GFS}(E)=\;\ln(|\bSigma_m|) $  \footnote{though $\pm m$ contribute to the same energy $E$, the domains of states  of $+m $ and $-m $ are within a single GFS not connected, i.e. with respect to ergodicity these two branches are separated. Hence, only one branch must be considered for entropy.}
 and inverse temperature $\kappa=\partial_E S_{GFS}(E) $ become with normalization of Boltzmann constant $k_{\text B}=1 $   
\begin{eqnarray}
S_{GFS}(E)&=&- \tfrac{1}{J N} E+S_{GFS}(0) \; \label{Eq:Entropy_singleGFS}\;,\\
\kappa &=& -\tfrac{1}{JN}\;.\label{Eq:TempGFS}\
\end{eqnarray} 
This means that the inverse temperature of a single GFS takes  almost constant negative values within the low energy regime, as denoted above (see also Appendix \ref{AppendixA0}).  

\subsection{Assembly of thermally weakly coupled GFSs}
Whereas the thermal interaction of a single GFS with another system was in detail considered previously \cite{bauer2022geometrically}, the focus of this manuscript lies on the characterization of a collective of thermally weakly coupled GFSs.  

The microstate of a group of $M$ GFSs, as introduced above, derives from microstates of the individual GFSs, $\boldsymbol{\sigma}=(\sigma_1,\hdots,\sigma_M)^T $. Weak mutual thermal coupling between the single GFSs makes the Hamiltonian of the group the sum of the individual Hamiltonians,  $H(\boldsymbol{\sigma})=\sum_{\nu=1}^{M} H^{(\nu)}(\sigma_\nu)$. The energy then depends on the magnetic moments $\bom= (m^{(1)},\hdots,m^{(M)})^T$ as 
\begin{equation}
E(\bom)=\frac{J}{2} \sum_{\nu=1}^{M} \left( m^{(\nu)}\right)^2=\frac{J}{2}\, \|\bom\|^2\label{Eq:Evsm}\;,
\end{equation} 
with the norm  $\|\bom\| =\sqrt{\bom^T\bom}$. The number of microstates with  magnetic moments,   $\bom$, derives from  Eqs.~(\ref{Eq:Energy},\ref{Eq:Evsm}) as 
\begin{equation}
|\bSigma_{\bom}| =\prod_{\nu=1}^{M}|\bSigma_{m^{(\nu)}}|=\left(\tfrac{1}{\sqrt{2\pi N}}\;2^{(N+1)}\right)^M\;\underbrace{e^{-\frac{1}{N}\;\frac{1}{2} \|\bom\|^2}}_{=e^{\kappa E(\bom)}}\;.\label{Eq:Sigma_EM}
\end{equation}
The number of realizations of magnetic moments with a certain energy $E$ is according to Eqs.~(\ref{Eq:Evsm}) proportional to the surface of the $M$-dimensional hypersphere with radius $m_M(E)=\sqrt{2E/J} $. With Eq.~(\ref{Eq:Sigma_EM}) the number of microstates with energy $E$ is (see appendix~\ref{AppendixA})
\begin{eqnarray}
|\bSigma_E |&=&  |\Sigma_{\|\bom\|=\sqrt{2E/J}}|\; c\; {\cal{S}}_M (2E/J)^{\frac{M-1}{2}}\label{Eq:Sigma_E0}\\
&=&\sqrt{2}\;\left(\tfrac{2^N}{\sqrt{\pi N}}\right)^M e^{\kappa E}\;{\cal{S}}_M \;\;(E/J)^{\frac{M-1}{2}}  \;,\label{Eq:Sigma_E}
 \end{eqnarray}  
with ${\cal{S}}_M=2\pi^{M/2}/\Gamma(M/2)$ as surface area of the $M$-dimensional unit sphere, $c $ some proportionality factor,  and $\Gamma$ as Gamma function. Boltzmann entropy is   
\begin{eqnarray}
S(E)&=&\ln|\Sigma_E|\cr
&=&\ln|\Sigma_{\|\bom\|=\sqrt{2E/J}}|+\ln\bigg(c{\cal{S}}_M \left(\tfrac{2E}{J}\right)^{\frac{M-1}{2}}\bigg)\label{Eq:Entropy10}\\
&=&\kappa\; E+\tfrac{M-1}{2}\ln(E/J)+C\; \cr\cr
&=&\kappa\; E+\tfrac{M-1}{2}\ln(-N\;\kappa E)+C\;,\label{Eq:Entropy1}
\end{eqnarray} 
with $C$ as the logarithm of the remaining, energy independent factors of Eq.~(\ref{Eq:Sigma_E}), which scales almost linearly with $M$ \footnote{Note that the terms related to the surface of the hypersphere in Eqs.~(\ref{Eq:Sigma_E0},\ref{Eq:Entropy10}) do not vanish in the ground state $E=0$ but become unity.}. The validity of this Gaussian based approximation with entropy obtained from direct counting of states and Stirlings approximation is shown in appendix~\ref{AppendixA}, and appendix~\ref{AppendixAb}, respectively. 
In contrast to normal systems with short range interaction, the entropy in Eq.~(\ref{Eq:Entropy1})  at energy $E$ does not emerge as the sum of entropies of its constituents, the GFSs, each at same energy $E/M$. Rewriting the above entropy of the assembly in terms of energy, $\varepsilon^{(\nu)}=1/2\; J (m^{(\nu)})^2 $, and entropy,  $S_{GFS}^{(\nu)}(\varepsilon^{(\nu)})$, of its constituents, i.e. the GFSs, demonstrates that instead one has  
\begin{eqnarray}
\Delta S(E)&=&\ln\left(\sum_{\substack{\varepsilon^{(1)},\hdots ,\varepsilon^{(M)}\\\sum \varepsilon^{(^\nu)}=E}}\exp\left(\sum_{\nu=1}^{M}\underbrace{\scriptstyle{\Delta S^{(\nu)}_{GFS}}(\varepsilon^{(\nu)})}_{=\kappa\varepsilon^{(\nu)}}\right)\right)\label{Eq:degtemperature1}\\ 
&=&\ln\left(\sum_{\substack{\varepsilon^{(1)},\hdots \varepsilon^{(M)}\\\sum \varepsilon^{(\nu)}=E}}\exp\left(\kappa\underbrace{\sum_{\nu=1}^{M}\varepsilon^{(\nu)}}_{=E}\right)\right)\label{Eq:degtemperature2}\\
&=&\kappa E+\ln\underbrace{\left(\sum_{\substack{\varepsilon^{(1)},\hdots \varepsilon^{(M)}\\\sum \varepsilon^{(\nu)}=E}}\;1\right)}_{\substack{\sim\text{surface of hypersphere}\\ \text{ with radius $\sim \sqrt{E} $}}} \cr\cr
&>& \kappa E=\sum_{\nu=1}^{M} \underbrace{\kappa\; E/M}_{\substack{\Delta\text{entropy of a single}\\ \text{  GFS at  $\varepsilon= E/M $ }}} \label{Eq:degtemperature3}\;.
\end{eqnarray}
Inequility~(\ref{Eq:degtemperature3}) implies that additivity of entropy does not hold here. In normal systems, this additivity results from the sharp peak which the probability of energy configurations exhibits around its maximum. This maximum is located at the configuration where energies of the constituents are all equal, i.e. in our case this would be $E/M$. This sharp peak of maximum probability implies that in normal systems this state is clearly dominating and stable. However, the degeneracy of temperature of a single GFS (Eq.~(\ref{Eq:degtemperature1})), i.e. the fact that all its energy states exhibit the same temperature, implies that any energy configuration $\varepsilon^{(1)},\hdots ,\varepsilon^{(M)} $ of $M$ thermally weakly coupled GFSs, which conserves overall energy,$\sum \varepsilon^{(\nu)}=E$, occurs with the same probability, $\sim e^{\kappa E}$ (Eq.~(\ref{Eq:degtemperature2})). Hence, one has not a stable dominating equilibrium state, but instead an indifferent equilibrium with the consequence of strong super additivity of entropy in Eqs.~(\ref{Eq:Entropy10},\ref{Eq:Entropy1},\ref{Eq:degtemperature3}). Note, the above approach may be generalized from  the Gaussian - to Stirling's  approximation of entropy, which holds beyond the low energy regime (appendix~\ref{AppendixAb}).  
 
The inverse temperature of $M$ thermally weakly coupled GFSs derives as     
\begin{equation}
\beta_M= \partial_E S(E)=\underbrace{\tfrac{1}{2} \; \bar{\varepsilon}^{\;-1}+\kappa}_{=\beta_{\infty}} - \tfrac{1}{2M}\bar{\varepsilon}^{\;-1}\; ,
 \label{Eq:beta_Ea}
\end{equation}
with $\bar{\varepsilon}=E/M$ as the specific energy, and $\beta_{\infty}$ as the asymptotic inverse temperature, which a very large number ($M\to\infty $) of weakly coupled GFSs has. The latter is sum of an inverse temperature resembling that of an ideal 1-D gas and that of a single GFS, $ \kappa$.   

Equation~(\ref{Eq:beta_Ea}) implies that, with reference to Boltzmann temperature, within a group of thermally weakly coupled GFSs, the single GFS is always hotter than the group as a whole, which is counter intuitive.  
At smaller energies, the inverse temperature of the group of GFSs is positive, $\beta_M>0 $. It monotonically decreases with increasing energy, eventually vanishes ($\beta_M \to 0$) at 
\begin{equation}
\bar{\varepsilon}_0=-\tfrac{M-1}{M} \frac{1}{2\kappa}\approx J\; \frac{N}{2}\;\text{for large $M$},\label{Eq:E0}
\end{equation}
and stays thereafter in the negative temperature regime. 
In this context it should also be mentioned that the above considerations about a single system being hotter than an assembly of its thermally weakly coupled copies would also hold if one switches the anti-ferromagnetic interaction to a ferromagnetic one, i.e. $J<0 $. In this case, however, all effects would be restricted to the positive temperature regime.

\subsection{Scaling properties and limiting cases}
In context with the above, it is worth to discuss some scaling properties of the GFS assembly.
The specific energy, i.e. average energy of $M$ GFSs, may be rewritten as 
\begin{eqnarray}
\bar{\varepsilon}&=&\frac{1}{M}\sum_{\nu=1}^{M}\frac{J}{2} \left(m^{(\nu)}\right)^2=
N\frac{JN}{2}\frac{1}{M}\sum_{\nu=1}^{M}\left(\frac{m^{(\nu)}}{N}\right)^2\cr\cr
&=& -\frac{N}{2\kappa}\; \overline{\mathfrak{m}^2}\;
\end{eqnarray}
where $\mathfrak{m}=m/N$ is the fraction of the magnetic moment related to the number of spins, i.e. the magnetization of a GFS. Insertion into Eq.~(\ref{Eq:beta_Ea}) then yields
\begin{equation}
\beta_M=\kappa \left(1-\left(1-\frac{1}{M}\right)\;\frac{1}{N\;\overline{\mathfrak{m}^2}}\right)\;.\label{Eq:scalebetavskappa}
\end{equation}
This implies that the difference of temperature of a GFS and its assembly is negligible if 
\begin{equation}
\overline{\mathfrak{m}^2}\gg \frac{1}{N}\; ,\label{Eq:ThermodynamicLimit1}
\end{equation}
i.e. vice versa it is relevant in the low energy regime, which we have in focus in this manuscript. With increasing magnetization/energy outside this regime the temperature of the GFS and its assembly converge. 

It should be mentioned that the way we formulate the spin interaction of the GFS on the geometry of a $N-1 $-dimensional simplex, which is analogue to the primitive elements forming spin ice \cite{moessner2006geometrical, moessner2001}, is not an extensive system, i.e. within a single GFS $\kappa$ is not an intensive parameter as it depends on the size of the system $\sim N^{-1}$. However this may be accomplished by renormalization of the interaction constant $J\to J'/N$,  an approach used to describe (anti)ferromagnetism \cite{campa2009statistical}. The magnetization $\mathfrak{m}=m/N$ is considered as an intensive parameter, making the single system Hamiltonian in Eq.~(\ref{Eq:H0}), $H=1/2\; J' \mathfrak{m}^2\;N $, and entropy in Eq.~(\ref{Eq:Entropy_singleGFS}) becomes, $\sim - 1/2\;\mathfrak{m}^2\; N $. Hence, by scaling with $N$ the latter both become extensive parameters, which makes temperature of the single system an intensive parameter $\kappa\to \kappa'=-1/J'$, and the thermodynamic limit  $N\to \infty $ exists. Because Eq.~(\ref{Eq:scalebetavskappa}) stays invariant under this rescaling, Eq.~(\ref{Eq:ThermodynamicLimit1}) predicts that $\beta_M'$ becomes identical with $\kappa'$ in this limit. In other words: for a constant, non-vanishing magnetization $\mathfrak{m}=m/N$, temperatures of assembly and constituents become equivalent in the thermodynamic limit. However, in this manuscript we stay at finite size GFSs for which the variance of magnetization $\overline{\mathfrak{m}^2} $ is within the magnitude of  $\frac{1}{N} $, i.e. the magnetic moment $\sqrt{\overline{m^2}}$ lies within the range of several standard deviations $\sqrt{N}$ of the Gaussian in Eq.~(\ref{Eq:Energy}). Here, temperature differences in Eq.~(\ref{Eq:scalebetavskappa}) are of relevance.     

\section{GFSs as canonical ensemble}
The in-equivalence of Boltzmann temperatures of an ensemble and its constituents requires reconsideration of the canonical ensemble, and the way its distribution is derived.   

In the microcanonical based version~\cite{baldovin2021statistical,puglisi2017temperature, cerino2015consistent},
the system and its bath, as two weakly coupled partners, form a microcanonical ensemble with energy $E$. The system is in some state $\sigma $ with energy $H(\sigma)=\varepsilon$, and it is considered to be small compared to its bath. The latter's entropy may then be approximated by $S_{bath}(E-\varepsilon)=S_{bath}(E)-\beta_{bath} \varepsilon $, with $\beta_{bath}=\partial_E S_{bath}(E)$ as its inverse Boltzmann temperature. Because  the bath itself may be built from many weakly coupled GFSs, Eq.~(\ref{Eq:beta_Ea}) implies  
\begin{equation}
\beta_{bath}=\beta_{\infty}\;.\label{Eq:bathensemble}
\end{equation}
The probability that the GFS is in the state $\sigma$ is given by the bath's entropy yielding the Boltzmann factor,
\begin{eqnarray}
p_{GFS}(\mathbf{\sigma})&\sim & e^{S_{bath}(E-\varepsilon)}\sim  e^{-\beta_{\infty}\; \varepsilon}\;.\label{Eq:Cano1}
\end{eqnarray}   

The information theory based ansatz focuses on the distribution function $\varpi(\sigma)$ of microstates $\sigma $ of a group of many weakly coupled GFSs. With $M$ GFSs this function underlies the constraint of energy conservation
\begin{equation}
g[\varpi]=\sum_{\sigma} \varpi(\sigma) H(\sigma)-E/M\overset{!}{=}0\;.\label{Eq:energyconservation}
\end{equation}
The probability to find  the distribution $\varpi(\sigma)$ is given by the Shannon entropy $S_{Sh}[\varpi]= -\sum_{\sigma} \varpi(\sigma)\ln(\varpi(\sigma))$  (see appendix~\ref{AppendixB}) as 
\begin{eqnarray}
w[\varpi]&\sim & \exp(M S_{Sh}(\varpi)) \;, \label{Eq:w(p)1}
\end{eqnarray}
i.e. its scales according to the large deviation principle \cite{touchette2009large} with the Shannon entropy as rate function. 
The most probable distribution is that with maximum Shannon entropy under the constraint of energy conservation, which is obtained with the Lagrange multiplier $\beta^{(c)}$ from 
\begin{equation}
\partial_{\varpi({\sigma})} \big(S_{Sh}[\varpi]- \beta^{(c)}g[\varpi]\big)=0 \label{Eq:LagrangeMP}
\end{equation}
as
\begin{equation}
\varpi_{max}(\sigma)\sim e^{-\beta^{(c)} H(\sigma)}\label{Eq:Cano2}\;.
\end{equation} 
The Lagrange multiplier $\beta^{(c)}$ is by definition the inverse canonical temperature. The advantage of the information theory based derivation is that the canonical temperature solely derives from properties of the single system, i.e. the GFS, and its specific energy $\bar{\varepsilon}=E/M$ in Eq.~(\ref{Eq:energyconservation}). Because the assembly of GFSs has as a bath many constituents ($M$ large), the large deviation scaling of the probability $w[\varpi]$ in Eq.~(\ref{Eq:w(p)1}) predicts that   it peaks sharply around its maximum at $\varpi_{max} $. Therefore the states of the assembly are almost solely distributed according to $\varpi_{max} $, which implies the identity  
\begin{equation}
\varpi_{max}(\sigma)=p_{GFS}(\sigma)\;. \label{Eq:Equivprobabilities}
\end{equation}
Consequently, the Boltzmann temperature of a bath, assembled from GFSs, is the same as the canonical temperature of a GFS, introduced as a Lagrange multiplier to maintain energy conservation, i.e. with Eqs.~(\ref{Eq:bathensemble},\ref{Eq:Cano1}) 
\begin{equation}
\beta^{(c)}=\beta_{\infty}=\beta_{bath}\;. \label{Eq:equivalence1}
\end{equation} 
In this context it should be noted that the Boltzmann temperature of a large set of $M'\gg 1$ GFSs within a bath of $ M\gg M'$ GFSs becomes identical with the canonical temperature $\beta^{(c)}$, in agreement with our usual notion of thermodynamics.  

Boltzmann and canonical temperature of a GFS differ, as Eqs.~(\ref{Eq:TempGFS}, \ref{Eq:beta_Ea}, \ref{Eq:equivalence1}) predict 
 \begin{equation}
\kappa <\beta^{(c)} \;.\label{Eq:betaBoltzmannCanonical}
\end{equation} 

This in-equivalence is related to the special equilibrium probability distribution of energy which a GFS exhibits in the canonical ensemble. The probability to find a GFS  with energy $\varepsilon $ in one branch
of magnetic moment, $m(\varepsilon)=\pm\sqrt{2J \varepsilon}$, \footnote{note that only one branch must be considered for determination of entropy because the domains with equivalent energy $\Sigma_m,\; \Sigma_{-m} $ are disjoint.} (see Eqs.~(\ref{Eq:Entropy_singleGFS}, \ref{Eq:TempGFS}, \ref{Eq:Cano1},  
 \ref{Eq:equivalence1})) is,  
\begin{eqnarray}
p_{GFS}(\varepsilon_{\pm})&=&\sum_{\substack{{\sigma| H(\sigma)=\varepsilon}\\{\sign(m(\sigma)=\pm 1}}} p_{GFS}(\sigma)\cr\cr%\sim e^{ -\beta^{(c)} \varepsilon}\;\underbrace{\sum_{\sigma| H(\sigma)=\varepsilon}1}_{=\exp(S_{GFS}(\varepsilon))}\cr\cr\cr%
&\sim & \exp(S_{GFS}(\varepsilon)-\beta^{(c)} \varepsilon)\cr\cr
&= &e^{S_{GFS}(0)}\; \exp((\kappa-\beta^{(c)})\varepsilon) \;.\label{Eq:maxprob}
\end{eqnarray}      
This equilibrium distribution is also confirmed by simulations of spin dynamics of a GFS in a heat bath (see appendix~\ref{AppendixC}). 
As $\kappa-\beta^{(c)}<0 $, the most probable state of a GFS is its degenerated ground state ($\varepsilon=0 $), and, with reference to energy, this is a global but not a local maximum, i.e. temperatures of GFSs $\kappa $ and bath $\beta^{(c)}$ differ here.
In contrast to a GFS, normal additive, i.e. extensive systems with a short ranged interaction, $(\bullet)_{ext} $, have a concave entropy function $S_{ext}(\varepsilon) $, the slopes of which, i.e. inverse temperatures, include  that of the bath. This implies the existence of a local maximum of $p_{ext}(\varepsilon)\sim \exp(S_{ext}(\varepsilon)-\beta^{(c)}\varepsilon)$ at some $\varepsilon^*(\beta^{(c)})$, i.e.
\begin{equation}
\underbrace{\partial_{\varepsilon} \left. S_{ext}(\varepsilon)\right|_{\varepsilon^*}}_{=\beta_{ext}}-\beta^{(c)}=0\;, \label{Eq:equivalence2}
\end{equation}
where canonical - and Boltzmann temperature become equivalent. This is independent from sign of temperature, and provoked the premature statement, that  Boltzmann and canonical temperature are always equivalent in the negative range as well \cite{buonsante2016dispute,abraham2017physics}, which was shown here not to hold in general. The existence of a local maximum of probability of energy at some  $\varepsilon^*(\beta^{(c)}) $ in normal systems  is the base for the Legendre transformation to the Massieu function  $S_{ext}\to \tilde{S}_{ext}(\beta^{(c)})=S_{ext}(\varepsilon^*(\beta^{(c)})-\beta^{(c)}\varepsilon^*(\beta^{(c)})=-\beta^{(c)} F(\beta^{(c)})$, with the Helmholtz free energy $F$ \cite{Swendson2017}. The inversion of this Legendre transformation reveals
the most probable energy from the Massieu function, $\varepsilon^{*}(\beta^{(c)})=-\partial_{\beta^{(c)}}\tilde{S}_{ext}(\beta{(c)})$, and the canonical entropy from free energy $S_{ext}(\beta^{(c)})=-\partial_{T^{(c)}} F$, with temperature $T=1/\beta$ \cite{matty2017comparison}.       

%The inequality of Boltzmann and canonical temperature of a GFS  also implies that it makes not sense to assign a GFS a Helmholtz free energy. Doing this formally by  Legendre transformation of  $S_{GFS}(\varepsilon) $, would reveal the Helmholtz free energy as
%% 
%\begin{equation}
%F(\beta^{(C)})=-\tfrac{1}{\beta^{(C)}} \left(\underbrace{S_{GFS}(\varepsilon^*(\beta^{(C)})) -\beta^{(C)} \varepsilon^*(\beta^{(C)})}_{=\sup_{\varepsilon>0}(S_{GFS}(\varepsilon) -\beta^{(C)} \varepsilon)}\right)\;. \label{Eq:Helmholtz}
%\end{equation}
%%
%However, this procedure would not transform towards the variable $\beta^{(C)}$ as the independent variable by which entropy could be determined from free energy as a potential. This would require that the supremum in Eq.~(\ref{Eq:Helmholtz}) is local maximum implying the equivalence of Boltzmann and canonical temperature as in Eq.~(\ref{Eq:equivalence2}), which is not the case.  

The in-equivalence of temperature of a single GFS and that of its bath consisting of many weakly thermally coupled GFSs (Eq.~(\ref{Eq:betaBoltzmannCanonical})) prohibits the existence of a local maximum of the probability to find a GFS at some energy, and by this the construction of a Legendre transformation or free energy in the usual way. However, the fluctuations above the ground state given by Eq.~(\ref{Eq:maxprob}) suggest to construct a partition function to derive the mean energy,  
\begin{eqnarray}
Z(\beta^{(c)})&=&\sum_{\sigma}\exp(-\beta^{(c)}\varepsilon(\sigma))\cr
&=&\sum_{m} \exp(S(\varepsilon(m))\exp(-\beta^{(c)}\varepsilon(m))\;, \label{Eq:Freeenergy1}
\\&& \text{and with Eq.~(\ref{Eq:Energy}) }\cr\cr
& \approx &\tfrac{1}{\sqrt{2\pi}}\;\frac{2^N}{\sqrt{N/4}} \sum_{m} \exp\big((\kappa-\beta^{(c)})\underbrace{\varepsilon (m)}_{=\tfrac{1}{2} J m^2} \big) \label{Eq:Freeenergy2}\;.
\end{eqnarray}  
For large $N$ and $\beta^{(c)}-\kappa <1 $ (see Appendix~(\ref{AppendixD})) the sum may be replaced by an integral in $ m$ over a Gaussian distribution,i.e.
% Integral over sum, $=\tfrac{1}{2} \sqrt{2\pi}\; \frac{1}{\sqrt{J(\beta^{(c)}-\kappa)}}$ as differential in integral is $\delta m=2 $ multiply with factor $\tfrac{1}{\sqrt{2\pi}}\;\frac{2^N}{\sqrt{N/4}} $ 
%%
\begin{equation}
Z(\beta^{(c)})=
\frac{2^N}{\sqrt{N}}\; \frac{1}{\sqrt{J(\beta^{(c)}-\kappa)}} \label{Eq:Freeenergy3}\;.
\end{equation}
Straightforward the cumulant generating function, 
\begin{equation}
\Phi(\beta^{(c)})=-\ln(Z(\beta^{(c)}))\;,
\end{equation}
determines with  Eqs.~(\ref{Eq:beta_Ea},\ref{Eq:equivalence1})  the mean energy
\begin{equation}
\langle \varepsilon\rangle_{GFS}=\partial_{\beta^{(C)}} \Phi(\beta^{(c)})=\tfrac{1}{2}\;\frac{1}{\beta^{(c)}-\kappa}=\bar{\varepsilon}\label{Eq:Energycumulant1}\;. 
\end{equation}
This self consistent expression is definitely not trivial. The mean energy $ \langle\varepsilon\rangle_{GFS}=J/2\langle m^2\rangle$  of a GFS within its ensemble results purely from energy fluctuations away from its most probable state, the degenerated ground state at $\varepsilon^*=0 $.  In contrast, in extensive systems with short ranged interaction, i.e. additive systems, mean energy and energy of the mode are almost identical \cite{swendsen2016negative}, even, if fluctuations play a role \cite{falcioni2011estimate}.   

Though the potential $\Phi$ reveals the mean energy and other moments from its cumulants, it has not the property of a Helmholtz free energy in the sense of a Legendre transformation, the inversion of which  would reveals the mode of energy.  As noted above this transformation requires that the mode of the energy is a local maximum, where temperature of the system and bath become equivalent. Then, by saddle point approximation, in the thermodynamic limit free energy and cumulant generating function would be directly related by $F(\beta^{(c)})\approx -1/\beta^{(c)}\Phi(\beta^{(c)})$, and mode and mean of energy would become equivalent, which is not the case here.

\section{Challenge for the 2nd law of thermodynamics}
\begin{figure}
\includegraphics[width=9.5cm] {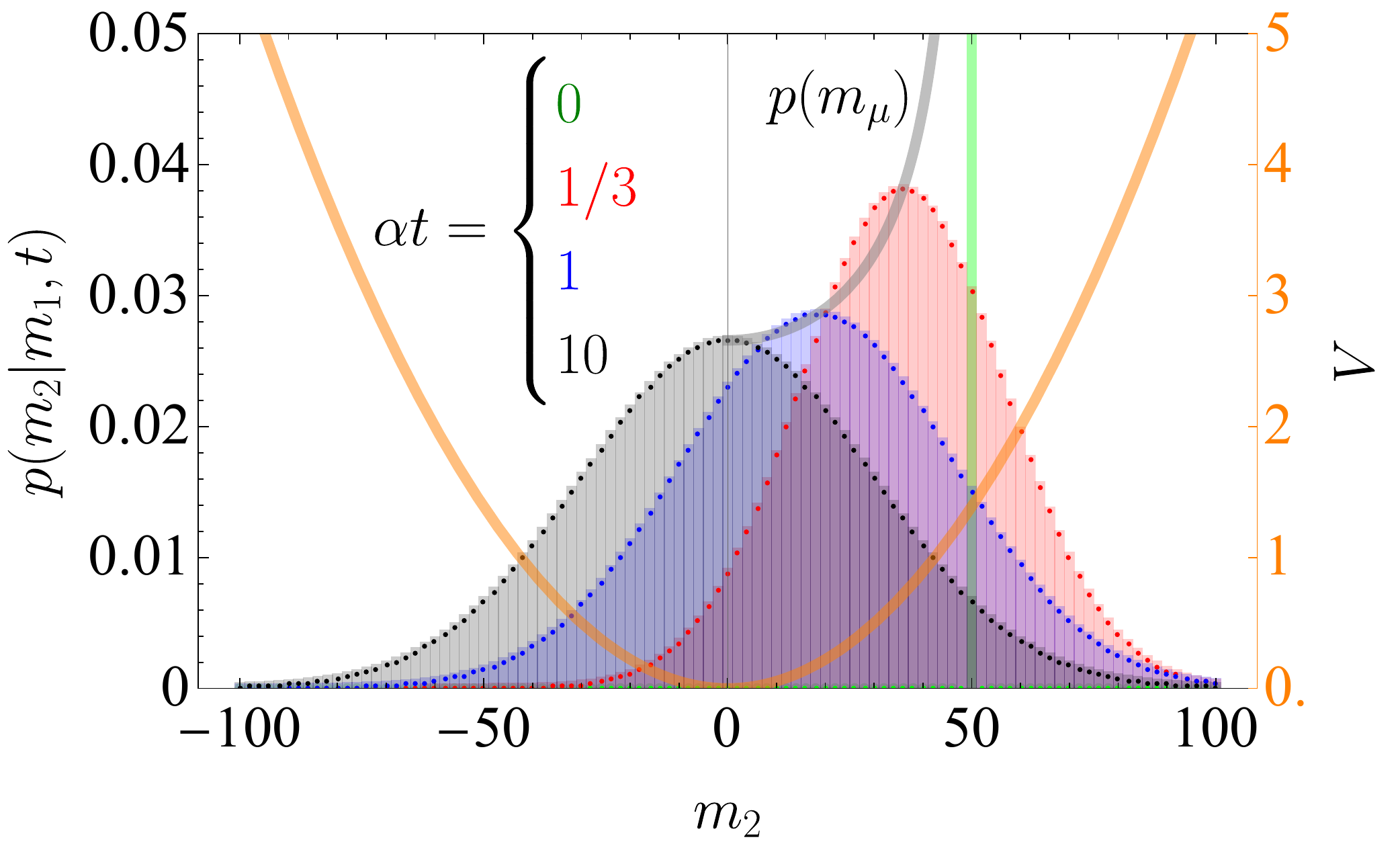}
\caption{Relaxation of the conditional probability distribution $p(m_2|m_1,t)$ (Eq.~(\ref{Eq:Greensfunction})), with initial magnetic moment $m_1=50$ (green) and three further subsequent points in time, $\alpha\; t$  ($\alpha=DJ(\beta^{(c)}-\kappa) $), color coded in red, blue and black. The GFS consists of $N=1000$ spins, i.e. temperature is $\kappa=-1/1000\;J^{-1}$. The canonical temperature $\beta^{(c)}$ was adjusted to a mean energy in the bath of $\langle\varepsilon\rangle_{GFS}=\tfrac{1}{2}\;(\beta^{(c)}-\kappa)^{-1}=\tfrac{J}{2}\langle m_2^2\rangle=\tfrac{J}{2}\; 30^2$. At $\alpha\;t=10$ the equilibrium distribution $p_{GFS}^{(m)}(m_2)=\lim_{t\to\infty}p(m_2|m_1,t)$  is almost reached. The harmonic potential $V $ affecting the transitions between  magnetic moments in orange (see text). In grey the evolution of the probability of the mode $p(m_\mu) $. Note: probabilities are given for  discretized magnetic moments.} 
\label{Fig:1}
\end{figure}
Referring to Boltzmann temperature a GFS is hotter than its bath assembled from many weakly coupled  GFSs (Eqs.~(\ref{Eq:equivalence1},\ref{Eq:betaBoltzmannCanonical})).  This provokes  the thought experiment:  a single GFS with energy $ \varepsilon_1$ taken out of its bath 1   and brought in contact for a time $t$ with another bath 2 at same temperature, should deliver heat according to the gradient of respective Boltzmann temperatures, i.e. $Q(t)=-\Delta\varepsilon(t)=-(\varepsilon_2(t)-\varepsilon_1)>0$, with $\varepsilon_2(t) $ as the GFS's energy in bath 2 after time $t $ . This is in particular intriguing, because prediction of heat flow from hot to cold is one of the main merits one normally assigns to  Boltzmann's temperature \cite{cerino2015consistent,PhysRevE.92.020103}. Obviously such a behavior would be in conflict with the 2nd law of thermodynamics as it would imply GFS mediated heat flow between two baths at the same temperature. Hence, the concept of heat flow from hot to cold defined by Boltzmann temperatures must be reconsidered. 

Indeed such heat transfer increases the combined system's entropy, $\Delta S_{GFS}(t)+\Delta S_{bath}(t)=-(\kappa-\beta^{(c)})Q(t)>0$ (Eqs.~(\ref{Eq:Cano1}, \ref{Eq:equivalence1}, \ref{Eq:maxprob})), i.e. the combined system is more likely to  drift towards lower energy of the GFS. Apparently repeating this procedure would imply unlimited  heat transfer.   
However, as we will show, fluctuations inherent to this procedure just compensate the effect of the the stochastic drift. 

Since elementary  stochastic spin flip processes directly translate into  changes of the magnetic moment, $|\delta m|=2 $, it is more convenient to switch to this state variable. 
The (equilibrium) probability to find a GFS in bath~1 with magnetization $m_1$ before taking it out is (see Eq.~(\ref{Eq:maxprob}) and spin dynamics simulations in appendix~\ref{AppendixC}) 
\begin{eqnarray}
p_{GFS}^{(m)}(m_1)&=&2\sqrt{\tfrac{J(\beta^c-\kappa)}{2\pi}}e^{-\tfrac{J}{2}(\beta^c-\kappa) m_1^2}
\;.\label{Eq:peqm}
\end{eqnarray} 
With $p(m_2|m_1,t)$ as conditional probability for the transition $m_1\to m_2$ after contact time $t$ in bath 2,  its sequential probability derives as $p(m_2|m_1,t)p_{GFS}^{(m)}(m_1)$. Exploiting conservation of probability, and detailed balance makes the energy balance for many repetitions of the above protocol vanish, since   
\begin{eqnarray}
\langle\Delta\varepsilon(t)\rangle_{m_2,m_1}&=&\langle \varepsilon_2(t)-\varepsilon_1\rangle_{m_2,m_1}=-\langle Q(t)\rangle\cr\cr
&=&\sum_{m_2, m_1}p(m_2|m_1,t) p_{GFS}^{(m)}(m_1)\tfrac{1}{2} J (m_2^2-m_1^2)\cr\cr
&= & \sum_{m_2} \tfrac{1}{2} J m_2^2 \underbrace{\sum_{m_1}p(m_2|m_1,t) p_{GFS}^{(m)}(m_1)}_{=p_{GFS}^{(m)}(m_2),\;{\text{detailed balance}}}\cr\cr\cr
&-&\sum_{m_1}\underbrace{\sum_{ m_2}p(m_2|m_1,t)}_{\substack{=1\\{\text{conservation of probability}}}} p_{GFS}^{(m)}(m_1)\tfrac{1}{2} J m_1^2 \cr\cr
&=&0 \label{Eq:energybil3}\;.
\end{eqnarray}   
Hence, though the single GFS of bath~1 is, by Boltzmann temperature, hotter than the assembly in bath~2  ($\kappa<\beta^{(c)}=\beta_{\text{bath 2}}$),  fluctuations emerging from the distribution $p_{GFS}^{(m)}(m_1) $ and the transition dynamics $p(m_2|m_1,t) $ impede net energy transfer between the baths, independent from the contact time $t$. Therefore one can state, that equal canonical temperatures predict absence of heat flow, despite a non-vanishing gradient of Boltzmann temperatures.

To illustrate this formal derivation we assume that the stochastic transitions between magnetic moments of a GFS in a bath result from Markovian spin flip processes. Then, in the continuum limit,  $\delta m\to 0$,  Equation~(\ref{Eq:peqm}) predicts dynamics of the magnetic moments as that of an Ornstein-Uhlenbeck process, i.e. diffusion within the harmonic potential $V(m)=\frac{1}{2}J(\beta^{(c)}-\kappa) m^2 $. The probability density \footnote{probability distributions on  discrete magnetic moments $p(m)\; (\delta m=2)$ and  density functions $\rho(m)\; (\delta m\to 0)$ in the continuum limit are related by $p(m)=\int_{m-1}^{m+1}dm'\rho(m')$ } evolves according to the Smoluchowski Equation (see ref.~\cite{Gardiner} and appendix~\ref{AppendixC}) 
\begin{equation}
\partial_t \rho(m,t)= D\;\partial_m\bigg(\partial_m-\underbrace{J(\kappa-\beta^{(c)})\; m}_{=-\partial_m V(m)}\bigg)\; \rho(m,t)\;,
\end{equation}    
with diffusion coefficient $D$. The conditional probability density $\rho(m_2|m_1, t) $, i.e. Greens function, becomes with the dynamic constant  $\alpha=D\;J (\beta^{(c)}-\kappa ) $  the Gaussian 
\begin{eqnarray}
\rho(m_2|m_1,t)&=&\tfrac{1}{\sqrt{2\pi}\sigma(t)}\exp\left(-\tfrac{(m_2-m_\mu(t))^2}{2 \sigma^2(t)}
 \right),\;\text{with}\cr\cr
m_\mu(t)&=& m_1 e^{-\alpha t},\;\text{and}\; \sigma^2(t)=\frac{1-e^{-2 \alpha t}}{J(\beta^{(c)}-\kappa)}
\label{Eq:Greensfunction}
\end{eqnarray} 
as mode and variance, respectively. Hence, the most frequent magnetic moment drifts from its initial position  towards the ground states, $m_\mu(0)=m_1\to 0$, (Fig.~(\ref{Fig:1}) and comparison of analytical results with spin-dynamics simulations in appendix~\ref{AppendixC}), i.e. parallel to the gradient of Boltzmann temperatures $\kappa<\beta_{bath}$ from the hot GFS ($\kappa $) to the cooler bath ($\beta_{bath}=\beta^{(c)}$, see Eq.~(\ref{Eq:equivalence1})). 
\begin{figure*}
\onecolumngrid
\centering
\includegraphics[width=16cm] {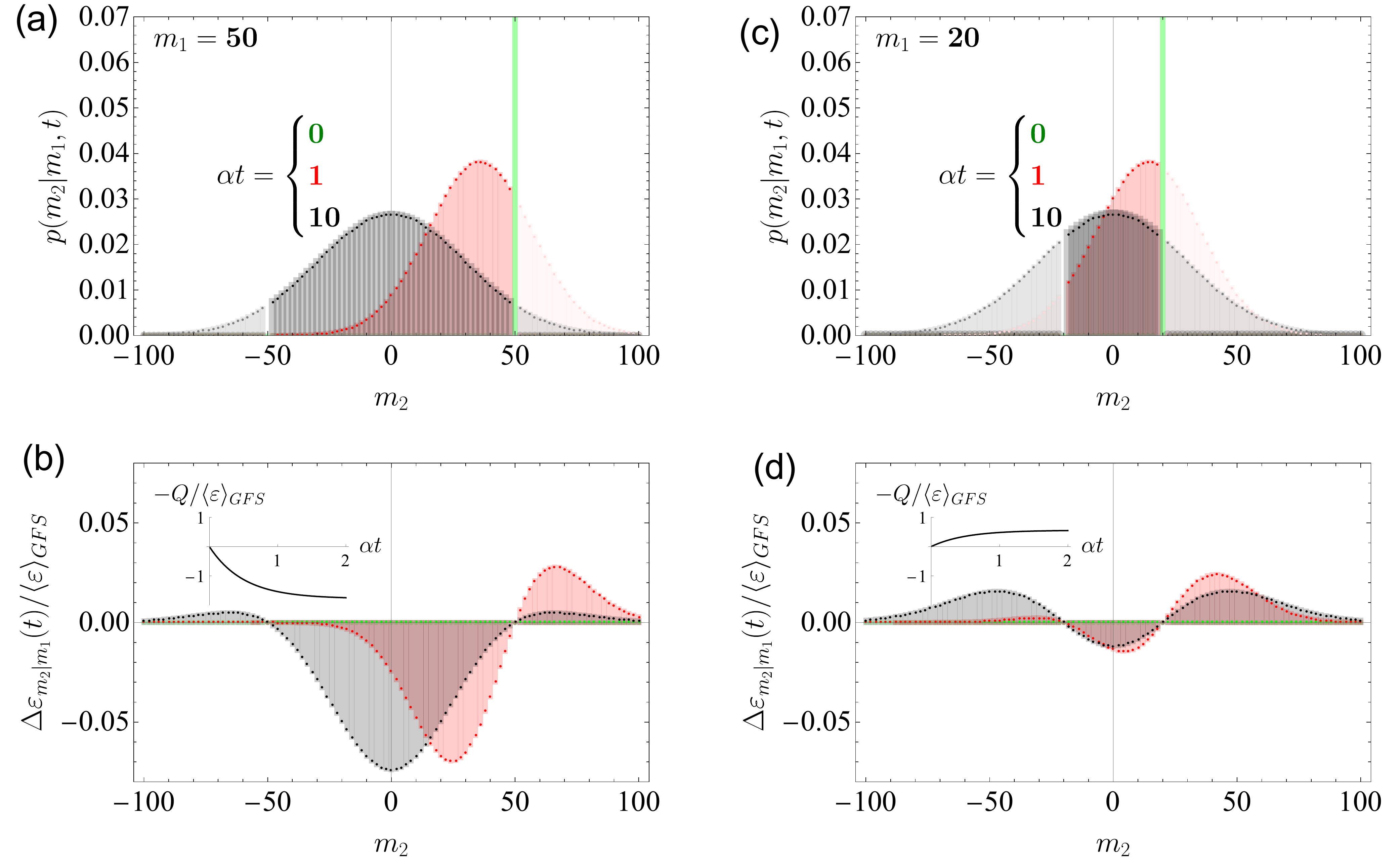}
\caption{Fluctuation mediated energy dissipation of a GFS in bath~2, visualized by relaxation of the conditional probability $p(m_2|m_1,t)$ (Eq.~(\ref{Eq:Greensfunction})) and corresponding energy balance $\Delta\varepsilon_{m_2|m_1}(t) $ (Eq.~(\ref{Eq:energybalance0})) for an initial moment  $m_1=50$ (a,b), and $m_1=20$ (c,d). Other parameters are as in Fig.~(\ref{Fig:1})). Black/gray and red/lightred in  (a,c) imply energy release/uptake of the GFS to /from the bath. According to Eq.~(\ref{Eq:energybalance1}) heat $ Q$ is released to bath~2 for $m_1=50$ (b) (negative areas under the curves and time course in insert) and taken up from bath~2 for $m_1=20$ in (d) (positive area under the curve and insert). } 
\label{Fig:2}
\twocolumngrid
\end{figure*}

The fraction of GFSs having undergone the transition $m_1\to m_2$ in bath~2,  contribute to the energy balance with 
\begin{equation}
\Delta\varepsilon (t)_{m_2|m_1}=\rho(m_2|m_1,t) \left(\tfrac{1}{2} J m_2^2 -\tfrac{1}{2} J m_1^2\right)\;.
\label{Eq:energybalance0}
\end{equation}
The impact of fluctuations from transition dynamics on energy dissipation, i.e. heat transfer from the GFS to bath 2, derives from  Eq.~(\ref{Eq:Greensfunction}) as conditional average
\begin{eqnarray}
Q(t) &=&-\int dm_2\; \Delta\varepsilon_{m_2|m_1} (t)=-\tfrac{J}{2}\left(\sigma^2(t)+m^2_\mu(t)-m_1^2 \right)\cr\cr
&=& \underbrace{-\tfrac{J}{2}\left(\tfrac{1}{J (\beta^{C}-\kappa)}-m_1^2\right)}_{=\varepsilon(m_1)-\langle\varepsilon\rangle_{GFS}}\left(1-e^{-2 \alpha t}\right)\;.\label{Eq:energybalance1}
\end{eqnarray} 
Therefore the gradient between initial energy of the particular GFS taken out from bath~1, $\varepsilon(m_1)$,  and the mean energy of GFSs in bath~2, $\langle\varepsilon\rangle_{GFS}$, (Fig.~(\ref{Fig:2})) is parallel to the direction of heat flow. Depending on the initial energy $\varepsilon(m_1)$, this flow may be parallel or anti-parallel to the gradient of Boltzmann temperatures.  

Additional effects of fluctuations originating from the equilibrium distribution in  bath~1 become evident by 
randomly picking GFSs out of it, followed by insertion in bath~2. Then average energy balance vanishes, 
\begin{equation}
\langle\Delta\varepsilon (t)\rangle_{m_2,m_1}=\int dm_2 dm_1 \;\Delta\varepsilon_{m_2|m_1} (t) =0\;,\label{Eq:energybalance2}
\end{equation}
i.e. no heat flows between baths at equal canonical temperatures. For different canonical temperatures $\beta^{(c)}_1\neq \beta^{(c)}_2$ the energy balance becomes  
\begin{eqnarray}
\langle\Delta\varepsilon (t)\rangle_{m_2,m_1}&=&\left(1-e^{-2 \alpha_2 t}\right) 
\tfrac{J}{2}\left(\tfrac{1}{J \beta^{(c)}_2-\kappa}-\tfrac{1}{J \beta^{(c)}_1-\kappa}\right)\cr\cr
&=&\left(\beta^{(c)}_1-\beta^{(c)}_2\right)
\tfrac{1-e^{-2 \alpha_2 t}}{2 (\beta^{(c)}_2-\kappa)(\beta^{(c)}_1-\kappa)}\;,
\end{eqnarray}
i.e. heat  $Q(t)=-\langle\Delta\varepsilon (t)\rangle_{m_2,m_1}$ flows parallel to the gradient of the canonical temperatures.

\section{Summary and conclusion}
GFSs exhibit constant negative temperature, which makes them an attractive paradigm to challenge our notion of thermodynamics.  In contrast to conventional paradigms for negative temperature, e.g. paramagnetic spin gases,  they are not subject for discussions referring to instability  \cite{ramsey1956thermodynamics,romero2013nonexistence,struchtrup2018work}. Neither do GFSs develop different phases, as systems in Ref.~\cite{ramirez2008systems}, which made them instable when brought in weak thermal contact. 

Negative temperatures were observed by numerical tests in spin ice \cite{raban2022violation}, i.e. realistic geometrically frustrated systems. However, they were obtained from analysis of fluctuation-dissipation relations as negative effective temperatures in non-equilibrium, glassy states. In our model, the situation is different, because we study the primitives of spin ice as the geometrically frustrated system, and not a lattice constructed from theses primitives. Because the GFS as a primitive is a finite system in which the inter-spin interaction is completely symmetrical, ergodicity holds and equilibrium is reached in finite time. Therefore, the concept of Boltzmann entropy and temperature, which is based on equal probability of microstates holds also for the GFS, i.e. the GFS has an equilibrium temperature.

A set of weakly coupled GFSs is, with reference to Boltzmann temperature, cooler than the single GFS itself, and may even  take positive temperatures. If this set consists of many GFSs, i.e. when this combined system may be considered as a heat bath, its Boltzmann temperature becomes equivalent with the canonical temperature of the GFS constituents, derived from an information technical approach. Hence, Boltzmann - and canonical temperature of a GFS differ. This in-equivalence  of Boltzmann and canonical temperature of a GFS also implies that a GFS cannot be assigned a Helmholtz free energy within its bath by a Legendre transform, as can be done for normal, i.e. additive systems.  Nevertheless, the associated partition function reveals correctly the mean energy of a GFS. 

The GFSs also differ from systems with long ranging interactions with ensemble in-equivalence \cite{baldovin2021statistical,ramirez2008systems,baldovin2018physical}, including also those showing this in the negative temperature domain \cite{caglioti1992special,smith1990nonaxisymmetric,miceli2019statistical}. This implies that fixing the energy (microcanonical ensemble) reveals for an observable a different result than fixing at the corresponding temperature (canonical ensemble). Often, a convex entropy function, i.e. negative specific heat, implies thermodynamic instability for a combined system and potentially coexistence of different phases. In contrast the constant, i.e. energy independent, temperature of the GFS imply an indifferent rather than unstable state of an assembly of GFSs, and different phases play no role.  
 
The difference of Boltzmann and canonical temperature of a GFS, i.e. the difference of Boltzmann temperatures of a single GFS and that of a bath consisting of many copies of this GFS,  leads apparently to a conflict with the 2nd law of thermodynamics, because,  at a first glance, this difference would imply heat transport. However, the Boltzmann temperature gradient predicts solely in which direction of energy uptake or release the most probables state, i.e. the mode,  of the combined system drifts. But fluctuations related to the stochastic dynamics of energy exchange and the equilibrium distribution of a GFS within its bath make heat flow parallel to the canonical temperature gradient, which may be opposed to that of Boltzmann temperatures. 

An intriguing aspect is that our concept of GFS is not restricted to physical systems in the classical sense. The GFS  used here, is a perfectly symmetrical network of interacting units, i.e. spins.
These spins could also be arranged within a neural network structure, e.g. Boltzmann machines.
Here,  temperature can be used as an adjusting parameter, which improves the performance of such learning machines \cite{li2016temperature}. Conceptually, this temperature parameter corresponds to our canonical temperature. The intrinsic temperature, i.e. the system's Boltzmann temperature, was not yet addressed. However, one has to admit, that it is a great step from our simple GFS with equivalent mutual interactions between all spins to neural networks in which the interaction between units is adjusted during learning. But the fact that even in the brain, as a biological network, the concept of frustration in the physical sense, emerges \cite{saberi2022pattern}, should be an impetus for further work in this field.

%\bibliographystyle{apsrev4-2}
%\bibliography{literature}
\bibliography{literature}

\begin{appendix}
\numberwithin{equation}{section} 
\renewcommand{\thesection}{\Alph{section}}
\setcounter{equation}{0}
\renewcommand\thefigure{\thesection.\arabic{figure}}
\setcounter{figure}{0}
\appendix\

\section{Exact and Gaussian based approximation of the Boltzmann temperature of a single GFS}
\label{AppendixA0}
\setcounter{figure}{0}
In the low energy regime ($m\ll N $) the Gaussian approximation of the number of microstates with magnetic moment $m$ of a single  GFS (Eq.~(\ref{Eq:Energy})) is the base for the analytical expressions of its Boltzmann temperature in the microcanonical ensemble, i.e. $\kappa $ (Eq.~(\ref{Eq:Energy})). The exact binomial distribution  yields a concave entropy vs. energy function, whereas the Gaussian approximation makes  entropy a linear function of energy. However, 
Figure (\ref{Fig:Suppl0}) shows that in the low energy range the Gaussian approximation of the number of microstates in Eq.~(\ref{Eq:Energy}) provides an excellent approximation of the inverse Boltzmann temperature, when compared with the result obtained from exact binomial distribution. The concave exact entropy function implies that its inverse temperature  is slightly smaller than the approximation $\kappa_{exact}<\kappa$, i.e. because temperatures are throughout negative, this implies $\kappa_{exact}/\kappa \gtrapprox 1$.   
\begin{figure}
\includegraphics[width=8cm] {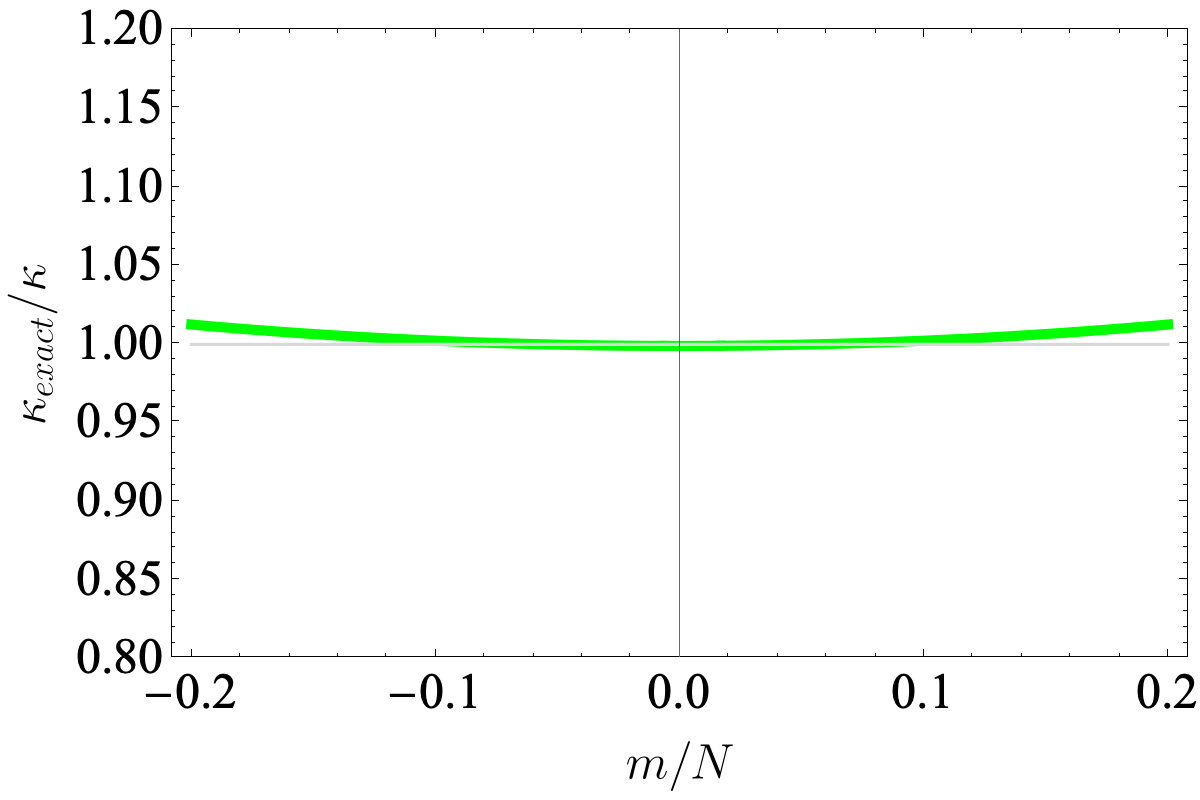}
\caption{Ratio of exact to Gaussian approximation of Boltzmann temperature of a GFS with $N=1000$ spins as a function of the magnetic moment $m$ . The exact inverse temperature  was obtained from the binomial distribution of the number of microstates for some magnetic moment $m$  (Eq.~(\ref{Eq:Energy})), i.e. $\kappa_{exact}=\partial_E \ln\binom{N}{\tfrac{N-m}{2}}$, with $E=1/2 J m^2$.  The   Gaussian approximation of the binomial reveals the inverse temperature $\kappa=-1/(JN)$ (see Eq.~(\ref{Eq:TempGFS})).} 
\label{Fig:Suppl0}
\end{figure}

\section{Number of states}\label{AppendixA}
\setcounter{figure}{0}
The number of microstates, which an assembly of $M$ GFSs with magnetic moments $\bom=(m^{(1)}),\hdots, m^{(M)})^T$ comprises, is according to Eq.~(\ref{Eq:Energy}) in the main text
\begin{eqnarray}
|\bSigma_{\bom}|&=&\prod_{i=1}^{M}\binom{N}{\tfrac{N+m^{(i)}}{2}}\;. \label{ASEq:Sigma_EM}
%% &\approx&\left(\tfrac{1}{\sqrt{2\pi N}}\;2^{(N+1)}\right)^M\;\underbrace{e^{-\frac{1}{N}\;\frac{1}{2} m_M^2}}_{=e^{\kappa E}}\;,\label{ASEq:Sigma_EM}
\end{eqnarray} 
The energy of this assembly is given by the Hamiltonian 
\begin{equation}
H(\bom)=\frac{J}{2}\; \sum_{i=1}^{M} (m^{(i)})^2=\frac{J}{2}\; \|\bom\|^2 \;.
\end{equation}
with the Euclidean norm of the magnetic moments $\|\bom\|=\sqrt{m^T m} $. The number of microstates  for some energy $E$  is then the sum over the number of states of magnetic moments $\bom $, which fulfill $H(\bom)=E $, i.e.  
\begin{equation}
|\bSigma_E|=\sum_{\bom} |\bSigma_{\bom}|\; \rect(2(\|\bom \|-\sqrt{2E/J}))\;. \label{AEq:NOSexact}
\end{equation}    
Here $\rect(x)$ denotes the boxcar function with $\rect(x)=1$ if $|x|\le 1/2 $, and otherwise zero, which takes into account the partitioning of the space of magnetic moments.\footnote{This counting of states within an interval of $\Delta m=2$ around some moment $\|m\|=\sqrt{2E/J}\pm1 $ is slightly different from the density of energy states, which is often taken as the base for determining Boltzmann entropy. This density would be obtained by counting states within some fixed energy interval. Our counting algorithm, which aims at the number of microstates for some energy $E=J/2\; \|m\|^2 $, exactly  reveals the correct number of states for $M=1$, i.e. the binomials of Ea.~(\ref{ASEq:Sigma_EM}), and is rigorously extended to $M>1$. Of note is that both approaches lead to the same result for larger $M$, as these relative difference of these differently defined entropies scales with $1/M $. } 

The binomials determining the number of microstates in Eq.~(\ref{ASEq:Sigma_EM}) may be approximated by Gaussians,
\begin{equation}
|\Sigma_{\bom}|\approx\left(\tfrac{1}{\sqrt{2\pi N}}\;2^{(N+1)}\right)^M\;\underbrace{e^{-\frac{1}{N}\;\frac{1}{2} \|\bom\|^2}}_{=e^{\kappa H(\bom)}}\;,\label{ASEq:Sigma_EM2}
\end{equation}  
which makes Eq.~(\ref{AEq:NOSexact}) 
\begin{equation}
|\bSigma_E|\approx \left(\tfrac{2^{(N+1)}}{\sqrt{2\pi N}}\right)^M e^{\kappa E}\sum_{\bom} \rect(2(\|\bom\|-\sqrt{2E/J})) \label{AEq:NOSgaussian}\;.
\end{equation}
The sum is
proportional to the surface ${\cal{A}}_M $  of the $M-$dimensional hypersphere with radius $m_M=\sqrt{2E/J} $, 
\begin{equation}
{\cal{A}}_M={\cal{S}}_M \;(2E/J)^{\frac{M-1}{2}} \; \label{AEq:shellsurface}
\end{equation} 
with 
  \begin{equation}
{\cal{S}}_M=2\pi^{M/2}/\Gamma(M/2) \;,
\end{equation} 
as the surface of the unit hypersphere. 
As magnetic moments are partitioned in units of 2 (one spin flip process), the $M-$dimensional space of magnetic moments is partitioned in cubic  voxels of volume  $2^M$. The magnetic moments with energy $E $ are then the voxels found within a spherical shell, the width of which is $2$ (see boxcar function in Eq.~(\ref{AEq:NOSexact}) and surface given by Eq.~(\ref{AEq:shellsurface}). This  surface itself is partitioned by the facet of the $M-$dimensional voxel, i.e. the $M-1 $ dimensional squares with sidelength $2$. Therefore the number of realizations of magnetic moments is 
\begin{eqnarray}
&&\sum_{\bom} \rect(\|\bom\|-\sqrt{2E/J})=\frac{{\cal{A}}_M}{2^{M-1}}\;,\cr\cr
&&={\cal{S}}_M\;(2E/J)^{\frac{M-1}{2}}\;2^{-(M-1)} \;.\label{AEq:numberofMS2}
\end{eqnarray}

Hence, with this Equation the number of microstates of the assembly of GFSs with energy $E$ in Eq.~(\ref{AEq:NOSexact}) becomes in the Gaussian approximation (Eq.~(\ref{AEq:NOSgaussian}))   
\begin{eqnarray}
|\bSigma_E |&\approx &{\cal{S}}_M\; (2E/J)^{\frac{M-1}{2}} 2^{-(M-1)}\left(\tfrac{1}{\sqrt{2\pi N}}\;2^{(N+1)}\right)^M\; e^{\kappa E}\cr\cr
&=&{\cal{S}}_M\; (E/J)^{\frac{M-1}{2}} \tfrac{2^{\frac{M-1}{2}}}{2^{(M-1)}}\left(\tfrac{1}{\sqrt{\pi N}}\;2^N \sqrt{2})\right)^M\; e^{\kappa E}\cr\cr
&=&{\cal{S}}_M  
\sqrt{2}\;\left(\tfrac{2^N}{\sqrt{\pi N}}\right)^M (E/J)^{\frac{M-1}{2}}  e^{\kappa E}\label{AEq:Sigma_E}\;.
 \end{eqnarray}  

%%
%with $\delta(a,b)$ as the coarse grained Kronecker function \footnote{This is conform with the  partitioning of magnetization, if  $\delta(a,b)=1 $, if $|a-b|\le1$, otherwise zero, i.e. it defines a range with magnitude two,  $a\pm 1 $, in which  it does not vanish.}.

\begin{figure}
\includegraphics[width=8cm] {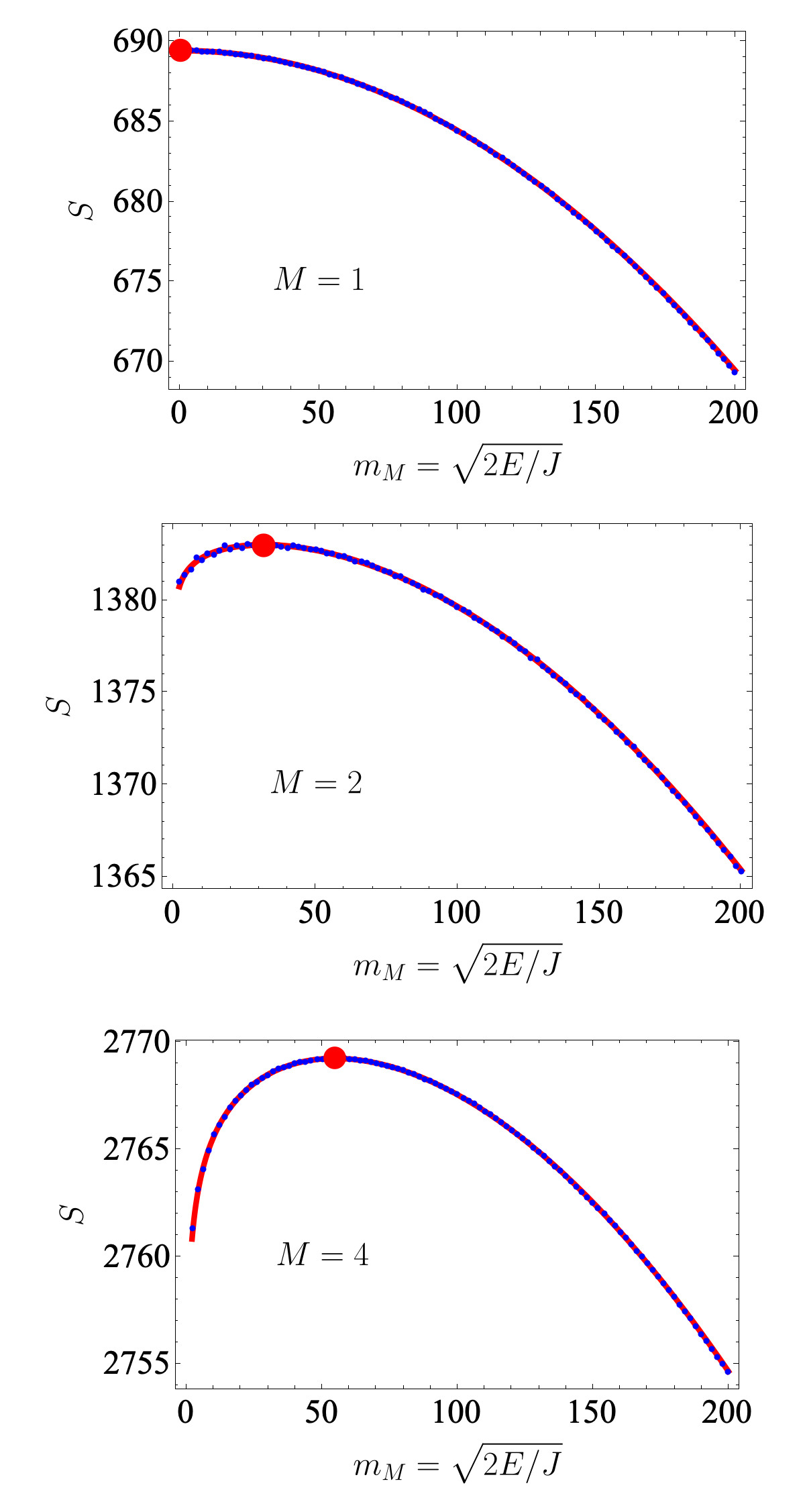}
\caption{Entropy $S=\ln(|\bSigma_E|$ vs. energy $E$, expressed by the norm of the magnetic moment $m_{M}=\sqrt{2E/J}$, of an assembly of $M=1,\;2,\;{\text{and}}\; 4$ GFSs, each consisting of $N=1000$ spins. Compared are the entropy obtained by counting the number of states from Eq.~(\ref{AEq:NOSexact}) (blue points) with the Gaussian based approximation in Eq.~(\ref{AEq:Sigma_E}) (red line). The energy where the inverse temperature $\beta_{M}=\partial_E S$  vanishes (see Eq.~(\ref{Eq:E0}) in main text), i.e. at $m_{M,\beta_M=0}=\sqrt{(M-1)N}\approx 0\;(M=1),\; 45\;(M=2),\;54\;(M=4)$  is shown by the red disk.}
\label{Fig:Suppl1}
\end{figure}

The quality of this Gaussian based approximation compared with direct counting of microstates in Eq.~(\ref{AEq:NOSexact}) is shown in Fig.~(\ref{Fig:Suppl1}). 

\section{Comparison of Gaussian with Stirling approximation of entropy}
\label{AppendixAb}
\setcounter{figure}{0}

\begin{figure*}
\onecolumngrid
\centering
\includegraphics[width=16cm] {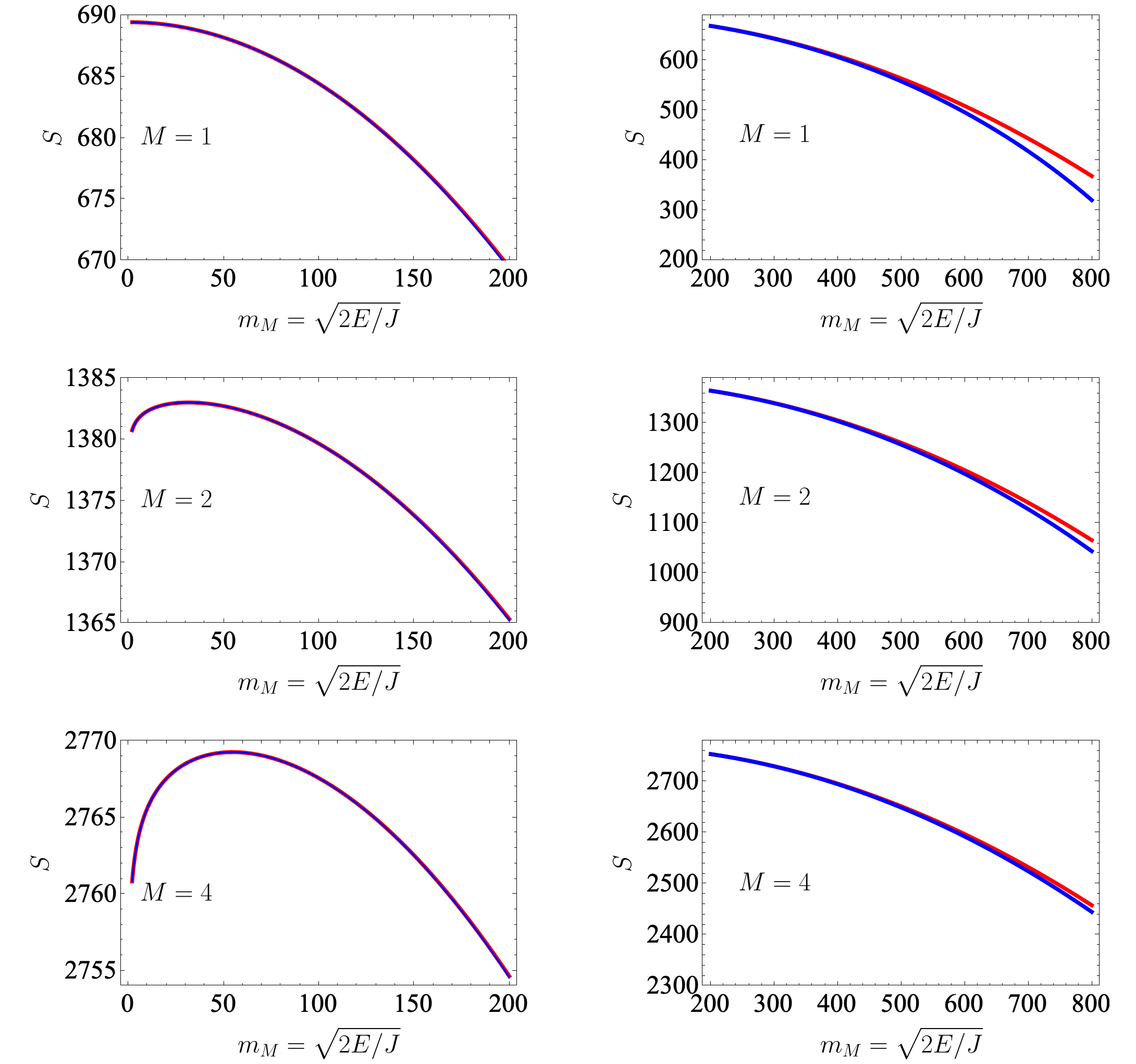}
\caption{Comparison of Stirling's (blue)  with Gaussian (red) approximation of entropy ($S=\ln(|\Sigma_E)|$), for the low and intermediate energy range. The number  $M$ of GFSs, and the setup of the individual GFS (number of spins $N$, coupling constant $J$) is that in Fig.~(\ref{Fig:Suppl1}). Energy is expressed by the norm of the moment $m_{M}=\sqrt{2E/J}$. The Gaussian approximation is derived from Eq.~(\ref{AEq:Sigma_E}). Stirling's approximation is given by $S(E)_{St}=\ln (|\Sigma_E|_{St})$ according of Eq.~(\ref{AEq:Stirlingnumberofstates1}), where the factor $f(m_M) $ is obtained from numerical integration in Eq.~(\ref{AEq:Stirlingnumberofstates2}). The low energy range is defined between $0< m_M< 200$, the intermediate within $200< m_M < 800$. }
\label{Fig:Suppl1b}
\twocolumngrid
\end{figure*}
\begin{figure}
\includegraphics[width=8cm] {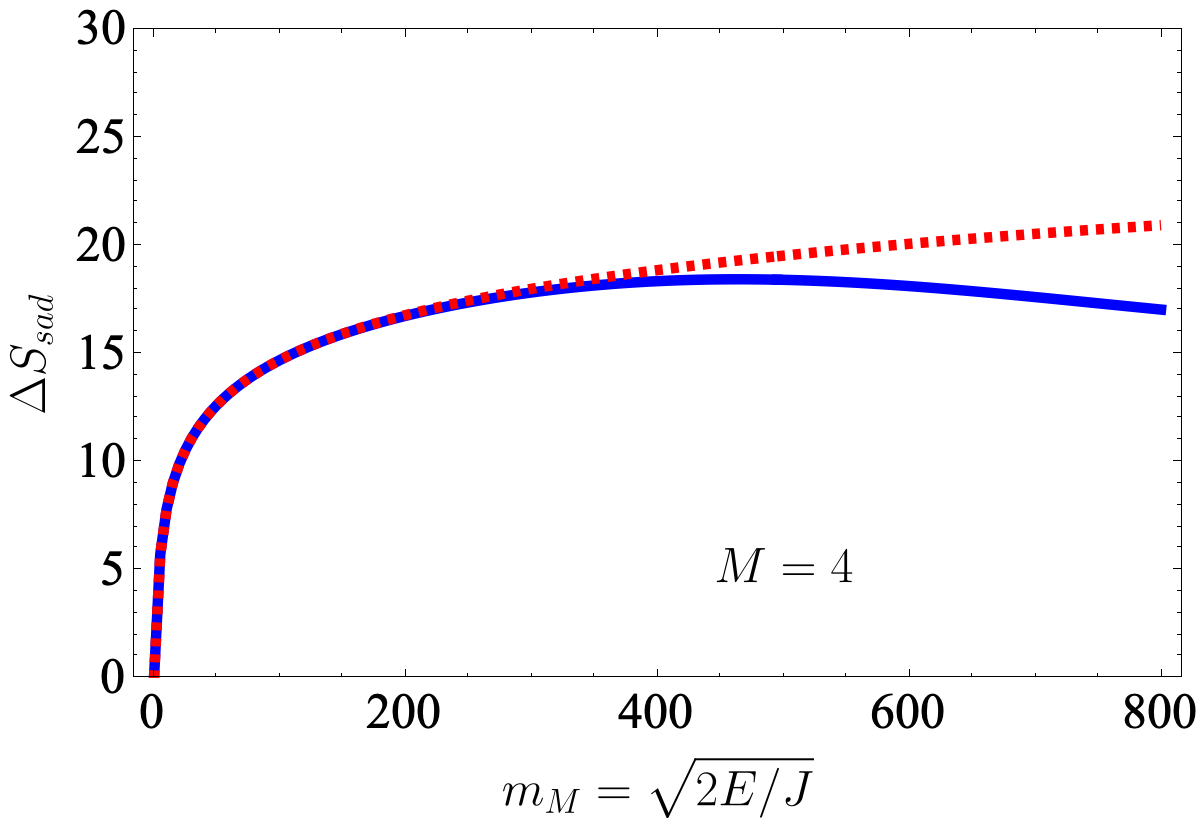}
\caption{Demonstration of superadditivity of entropy in the Stirling (blue) and Gaussian (red) approximation of an assembly of $M=4$ GFSs according to Eq.~(\ref{AEq:superadd}).  Stirling's approximation of entropies derives from  Eq.~(\ref{AEq:Stirlingnumberofstates1}), $\ln (|\Sigma_E|_{St}|)$; the Gaussian approximations  from Eq.~(\ref{Eq:Entropy1}) for the assembly and Eq.~(\ref{Eq:Entropy_singleGFS}) for the single GFS. Note that both curves almost coincide until $m_M=400$. }
\label{Fig:Suppl1c}
\end{figure}

Stirling's approximation of entropy based on binomials $\binom{N}{\tfrac{N+m}{2}}$ is considered to be more robust than the Gaussian one (Eq.~(\ref{Eq:Energy})), in particular for larger values of magnetic moments $m$. We show here, that even for intermediate large magnetic moments, the Gaussian approximation holds well.
In Stirling's approximation one has with $x=m/N$
\begin{eqnarray}
\ln \binom{N}{\tfrac{N+m}{2}}&\approx& N\ln(2)-\cr\cr
\tfrac{N}{2}((1+x)&\ln&(1+x)+(1-x)\ln(1-x))\;. 
\end{eqnarray}
The Gaussian approximation derives from  Stirling's approximation  by series expansion of the latter in $x$ to the second order, i.e. $\tfrac{N}{2}((1+x)\ln(1+x)+(1-x)\ln(1-x))\approx -N/2\;x^2 $. The difference between both approximation is the correction term
\begin{equation}
C(x)=-\tfrac{N}{2}((1+x)\ln(1+x)+(1-x)\ln(1-x)-x^2)\;.
\end{equation}
For an assembly of $M$ GFSs, the number of states with moments $\bom=(m^{(1)},\hdots, m^{(M)})^T $ in  Stirling's approximation $(\bullet)_{St} $  may be written
\begin{equation}
|\Sigma_{\bom}|_{St}=|\Sigma_{\bom}|_{G}\exp\left(\sum_{i=1}^{M} C\left(\tfrac{m^{(i)}}{N}\right)\right)\;,
\end{equation}
with $|\Sigma_{\bom}|_{G}$ as the Gaussian approximation (Eq.~(\ref{Eq:Sigma_E0})). The number of states with some energy $E=1/2\;J \|\bom\|^2)$ becomes
\begin{eqnarray}
|\Sigma_E|_{St}&=&|\Sigma_E|_{G}\tfrac{1}{{\cal{A}}_M}\int dm^{(1)}\hdots dm^{(M)}\delta(\|\bom\|-\sqrt{2E/J})\cr
&\times & \exp\left(\sum_{i=1}^{M} C\left(\tfrac{m^{(i)}}{N}\right)\right)\cr\cr\cr
&=&|\Sigma_E|_{G}\;f^{(M)}(m_M)\;,\label{AEq:Stirlingnumberofstates1}
\end{eqnarray}
with ${\cal{A}}_M$ as the surface of the $M$-dim hypersphere with radius $m_M=\sqrt{2E/J}$  and the factor  
\begin{eqnarray}
f^{(M)}(m_M)&=&\tfrac{1}{{\cal{S}}_M}\int_{\text{surface unit sphere}} d{{\cal{S}}_M}(\varphi^{(1)}\hdots d\varphi^{(M-1)})\cr
&\times & \exp\left(\sum_{i=1}^{M} C\left(\frac{m_M g^{(i)}(\varphi^{(1)}),\hdots,\varphi^{(M-1)})}{N}\right)\right)\;.\cr \label{AEq:Stirlingnumberofstates2}
\end{eqnarray} 
Note that $f^{(1)}(m_M)=\exp(C(m_M/N)$. 

The angles $\varphi^{(i)}$ form together with the radius $m_M$ hyperspherical coordinates in $M$-dimensions;   ${\cal{S}}_M={\cal{A}}_M/m_M^{M-1}$ is the surface of the $M-$dim unit sphere, and $d{{\cal{S}}_M}$ its surface element. The trigonometric functions $g^{(i)}$ link the Cartesian coordinates $m^{(i)}$ with the hyperspherical coordinates $m^{(i)}=m_M\; g^{(i)}(\varphi^{(1)}),\hdots,\varphi^{(M-1)})$.
Equations~(\ref{AEq:Stirlingnumberofstates1},\ref{AEq:Stirlingnumberofstates2}) derive the number of states in Stirling's approximation from that in the Gaussian one by multiplication with the factor $f$ (Eq.~(\ref{AEq:Stirlingnumberofstates2})). Accordingly the entropies in the respective approximations differ by the summand $\ln(f^{(M)}(m_M))$. Figure~(\ref{Fig:Suppl1b}) shows the perfect congruence in the low energy range ($0 < m_M < 200 $), but also in the intermediate regime ($ 200 < m_M < 800 $) both approximations provide similar results.  

The Stirling approximation of entropy also demonstrates that the strong superadditivity of entropy of an assembly of GFS, which becomes strikingly evident due to degeneracy of temperature in the Gaussian approximation (Eqs.~(\ref{Eq:degtemperature1}-\ref{Eq:degtemperature3})), holds for a wide range of energies. Note, that strong superadditivity  in the low energy regime  makes positive temperatures of the assembly emerge out of  constituents (GFSs), living at negative temperatures. Entropic superadditivity is defined as the difference of the entropy of the assembly of $M$-GFSs versus the sum of entropies of its constituents all at mean energy $E/M$, 
\begin{eqnarray}
\Delta S_{sad}(E)&=&S^{(M)}(E)-\sum_{\nu=1}^{M} S_{GFS}(E/M)\cr
&=&S^{(M)}(E)-M S_{GFS}(E/M)\label{AEq:superadd}\;.
\end{eqnarray}
From Eqs.~(\ref{AEq:Stirlingnumberofstates1}, \ref{AEq:Stirlingnumberofstates2}) follows that the superadditivi ty of entropies in the respective approximations, i.e. Stirling's ($St $) and Gaussian approach ($G$), is related by 
\begin{eqnarray}
\Delta S_{sad,\;St}(m_M)=\Delta S_{sad,\;G}(m_M)&+&\cr\cr
\ln(f^{(M)}(m_M)-M\ln\left(f^{(1)}\left(\tfrac{m_M}{\sqrt{M}}\right)\right)&& \label{AEq:superaddStvsG}\;.
\end{eqnarray}
Note, that if energy is measured by the radius of the hypersphere, $E\to m_M=\sqrt{2E/J}$,  the mean energy of the single GFS  transforms as $E/M\to m_M/\sqrt{M}$.
Figure~(\ref{Fig:Suppl1c}) demonstrates that the Gaussian approximation shows excellent agreement in predicting superadditivity in the low energy regime ($0 < m_M < 200 $) and even beyond. From $m_M>400$ the Gaussian approximation starts to overestimate superadditivity.   
    
\section{Canonical distribution derived from information theory}\label{AppendixB}
The information theory related ansatz focuses on how microstates $\sigma $  are distributed within a large assembly of GFSs. The microstate of the latter derives from the individual microstates  of its $M$ constituents, $\sigma^{(i)},\;i=1,\hdots, M$,  as 
$\boldsymbol{\sigma}= (\sigma^{(1)},\hdots,\sigma^{(M)})^t$, under the constraint of energy conservation, $\sum_{i=1}^{M} H^{(i)}(\sigma^{(i)})=E $.  The relative distribution frequency of the microstates is  
\begin{eqnarray}
\varpi(\sigma)&=&n(\sigma)/M \;, \text{with}\\
n(\sigma)&=&\sum_{i=1}^{M}\delta_{\sigma,\sigma^{(i)}}\;,
\end{eqnarray}
as the absolute frequencies. 

The degeneracy of a given distribution $\varpi(\sigma) $, i,e, the number of microstates $\boldsymbol{\sigma}$ which distribute accordingly, derives from  combinatorics as 
\begin{equation}
{\cal{D}}[\varpi]=M!/\prod_{\sigma} n(\sigma)! \;.\label{AEq:combinatorics}
\end{equation}
The logarithm of this degeneracy is proportional to the information one gains from knowing the distribution to knowing its micro state variable, $\varpi(\sigma)\to \boldsymbol{\sigma}$, i.e. knowing the distribution is less informative than the knowledge of the microstate.   
As equipartition predicts equal probability of all microstates $\boldsymbol{\sigma} $, which fulfill  energy conservation,  the degeneracy of a distribution is proportional to the probability of its occurence, $w[\varpi] $. Applying Stirling formula to Eq.~(\ref{AEq:combinatorics}), yields the probability to find  the distribution $\varpi(\sigma) $ as  
\begin{eqnarray}
w[\varpi]&\sim &  {\cal{D}}[\varpi]=\exp(M S_{Sh})\;,\;\text{with}  \label{AEq:w(p)1}\\ \label{Eq:w(p)2}
S_{Sh}[\varpi]&=& -\sum_{\sigma} \varpi(\sigma)\ln(\varpi(\sigma))\;
\end{eqnarray}
as Shannon entropy. 

%\section{Extensivity of Entropy of an Ensemble of GFS}
%The entropy of an ensemble of weakly coupled GFSs derives from Eq.~(\ref{AEq:Sigma_E}) for large number of %ensemble members ($1/M$ small) with the help of Stirling formula as
%\begin{eqnarray}
%S(E)&=&\ln(|\bSigma_E |)\cr\cr
%&=&M\left[\frac{1}{2}\ln\left(2 \bar{\varepsilon}/J \right)+\kappa\bar{\varepsilon} + C \right]
%\end{eqnarray}
%where $\bar{\varepsilon}=E/M $ is the specific energy, and  $C$ is some constant, independent of $M $ and $E$ . Hence, on the coarse grained scale of GFSs the Boltzmann entropy of an ensemble is extensive and the thermodynamic limit holds.  

\section{Binomial versus Gaussian approximation for determination of the cumulant generating function}\label{AppendixD}
\setcounter{figure}{0}
\begin{figure*}
\onecolumngrid
\centering
\includegraphics[width=16cm] {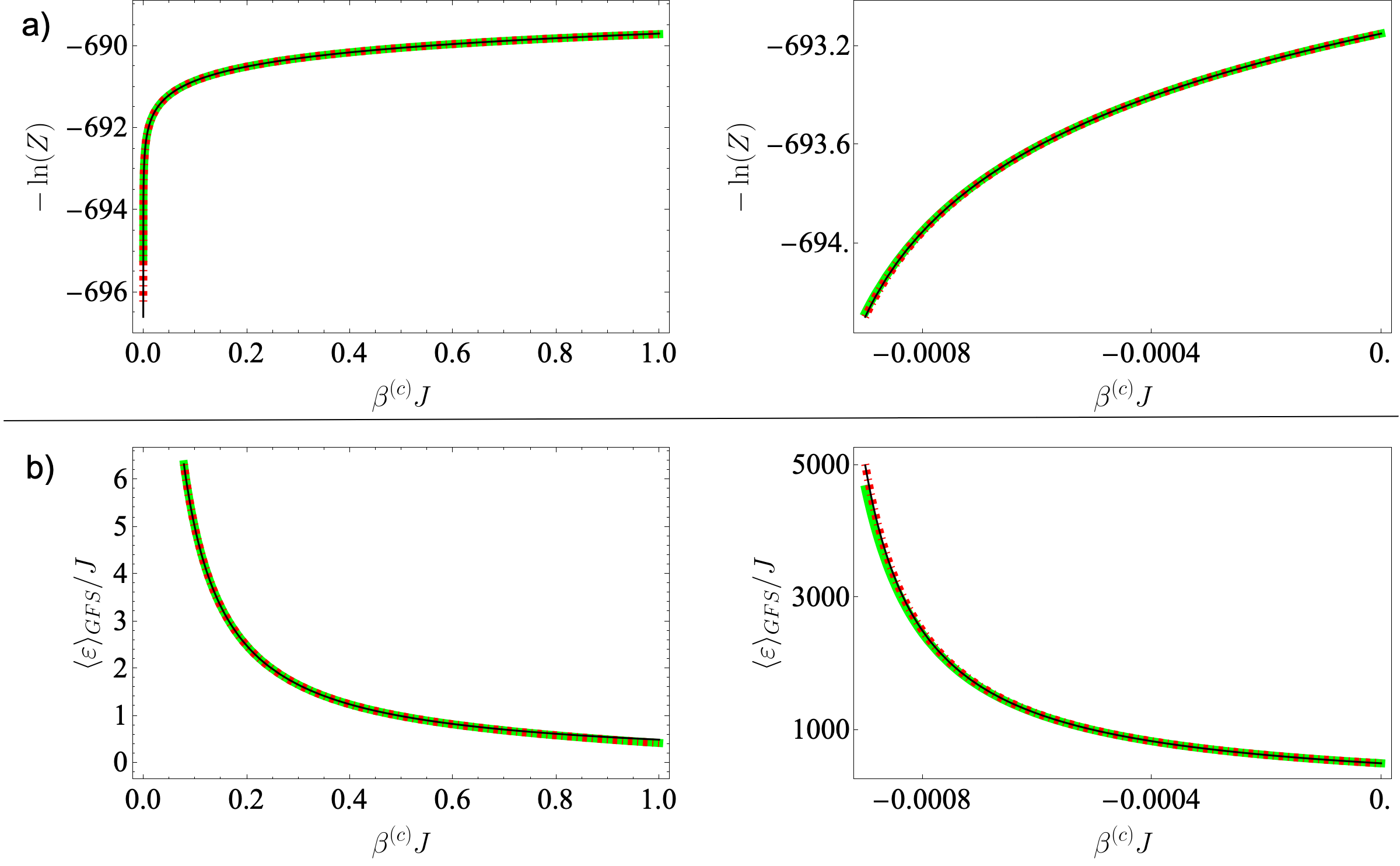}
\caption{Cumulant generating functions, $\Phi(\beta^{(c)}=-\ln(Z(\beta^{(c)})) $, with partition functions $Z$ (a)  and mean energy $\langle \varepsilon\rangle_{GFS}$ (b) of a GFS vs. inverse canonical temperature $\beta^{(c)}$ of the bath. The right chart in each row is a magnified section of the left for small inverse temperatures. The partition function is obtained from Eqs.~(\ref{Eq:Freeenergy1}) either, - with the exact binomial expression of entropy, $\exp(S(\varepsilon(m))=|\Sigma_m|=\binom{N}{\tfrac{N+m}{2}}$ of Eq.~(\ref{Eq:Energy}) (green), - its Gaussian approximation in  Eq.~(\ref{Eq:Freeenergy2}) (red dotted), - and the integral approximation of the latter sum in Eq.~(\ref{Eq:Freeenergy3}) (black). Note that the integral approximation implies $\sum_m f(m)\to 1/2 \int dm f(m) $ as the magnetic moment is partitioned  in units of $\delta m=2$.  As in the previous examples the number of spins of a GFS is $N=1000$, and hence its inverse temperature $\kappa=-1/1000 J^{-1}$.  Mean energies were either obtained by numerical differentiation $\langle\varepsilon\rangle_{GFS}=\partial_{\beta^{(c)}}\Phi=- \partial_{\beta^{(c)}}\ln(Z(\beta^{(c)}))$ for the binomial and Gaussian partition sum, or by Eq.~(\ref{Eq:Energycumulant1}) for the integral approximation of the Gaussian sum.  } 
\label{Fig:Suppl5}
\twocolumngrid
\end{figure*}

The validity of the Gaussian approximation of the exact, binomial distributed number of microstates for a magnetic moment, i.e.  $\exp(S(m))\sim e^{\kappa \tfrac{1}{2}m^2}$ vs. $\exp(S(m))=\binom{N}{\tfrac{N-m}{2}} $ is tested for determination of the partition sum (Eq.~(\ref{Eq:Freeenergy1})), the respective cumulant generating function and the thereof derived mean energy. In addition the integral approximation of the Gaussian partition sum in Eq.~(\ref{Eq:Freeenergy3}) is investigated. Figure~(\ref{Fig:Suppl5}) shows congruent curves of the exact (green) functions, its Gaussian approximation (red dotted) and the integral approximation of the latter (black). In particular, the relation between canonical inverse temperature and mean energy in Eq.~(\ref{Eq:Energycumulant1}), which is based on the Gaussian approximation, is excellently retrieved.  
%For very low temperatures of the bath, $\beta^{(c)}\to + \infty$, i.e. when the GFSs are almost in the ground state $m\; \text{and}\; \langle^\varepsilon\rangle_{GFS}\to 0 $ the integral approximation may fail, as $Z_exact\approx \exp(S(0)) e^{-0 }+ otherwise zero= \binom{N}{N/2}$ whereas the integral approximation gives $Z\to 0 $  

\section{Simulation of spin dynamics of a single GFS}\label{AppendixC}
\setcounter{figure}{0}
\subsection{General aspects and determination of equilibrium probability of a GFS within a heat bath}
\begin{figure}
\includegraphics[width=8cm] {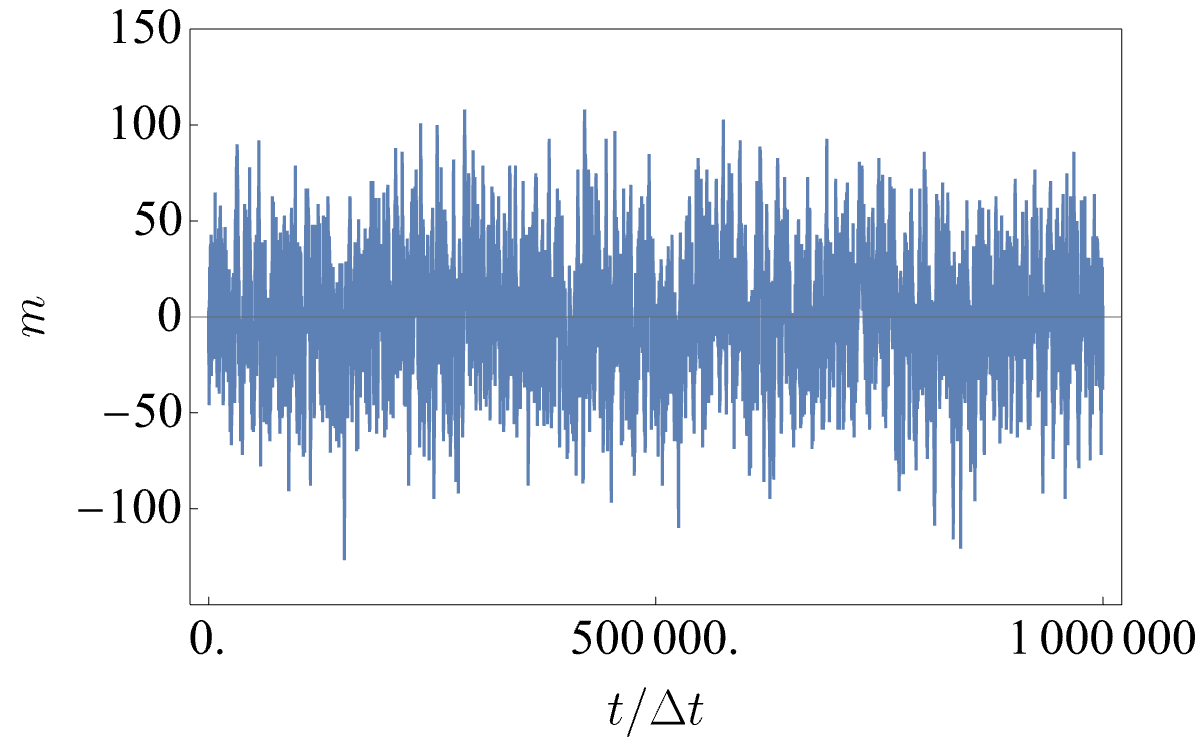}
\caption{Simulated stochastic trajectory in the space of magnetic moments $m $  for $t/\Delta t=10^6$ time steps. Starting point was at $m(0)=0$. Like in the example in the main text the GFS consists of $N=1000$ spins, i.e. its inverse temperature is $\kappa=-1/1000\;J^{-1}$, and the inverse temperature of the bath $\beta_{bath}$ was adjusted to a value that the mean energy of the GFS within is $\langle\varepsilon\rangle_{GFS}=\tfrac{1}{2}\;(\beta_{bath}-\kappa)^{-1}=\tfrac{J}{2}\langle m_2^2\rangle=\tfrac{J}{2}\; 30^2$.} 
\label{Fig:Suppl2}
\end{figure}
\begin{figure}
\includegraphics[width=8cm] {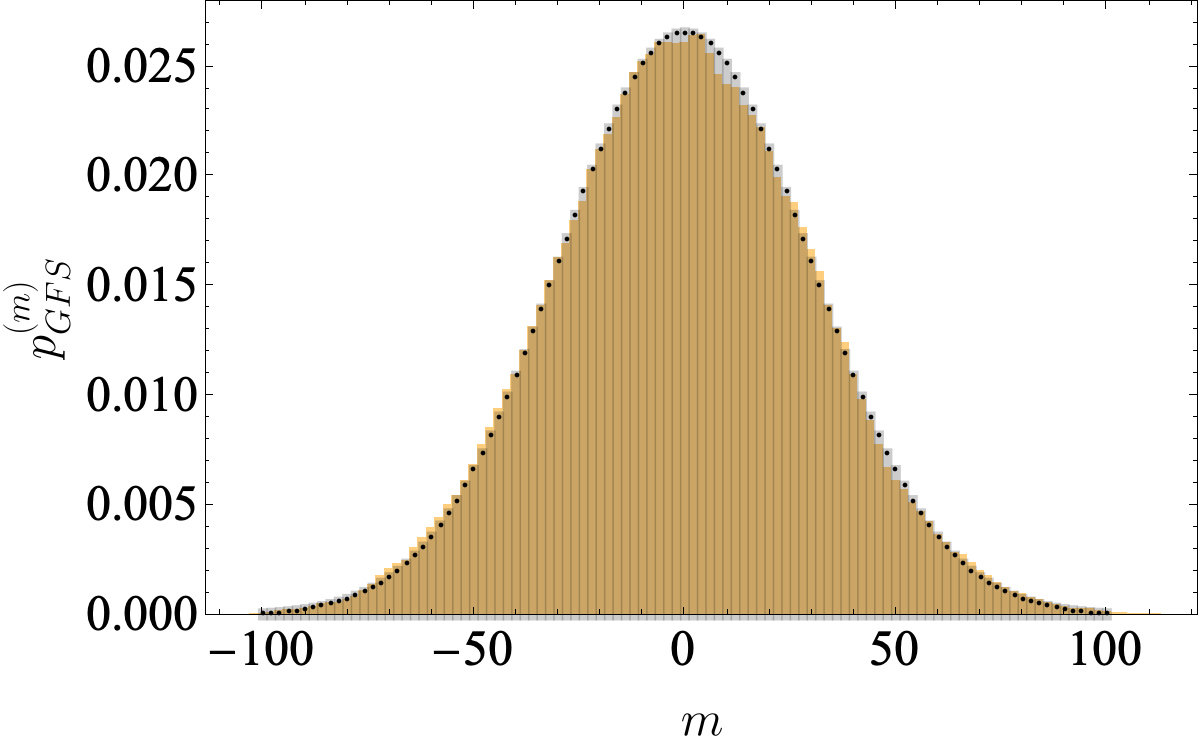}
\caption{Comparison between equilibrium distribution $p^{(m)}_{GFS}$ of the magnetic moment obtained  from the simulated stochastic trajectory in Fig.~(\ref{Fig:Suppl2}) (relative frequency of visits in brown histogram) and the canonical distribution (dotted line) obtained from $\exp(-1/2 \;J(\beta^{(c)}-\kappa) m^2)$ by discretization of magnetic moments in intervals of $|\delta m|=2 $ (see main text).  Note that canonical and bath temperature are equivalent, $\beta_{bath}=\beta^{(c)} $  (see main text). } 
\label{Fig:Suppl3}
\end{figure}

We assume that the microstate of a GFS $\sigma=(s_1,\hdots, s_N)^T$ with spin orientations $s_i=\pm1$ is subjected to stochastic forces originating from the heat bath with inverse temperature $\beta_{bath}$. These forces induce with some probability $p $ after some time $\Delta t $ the flipping of a single arbitrary spin $s_i\to -s_i$. The dipole moment $m=\sum_i s_i $ changes by  $\delta m=-2 s_i$, and energy 
\begin{eqnarray}
\Delta E &=& \tfrac{1}{2} J\;\delta m^2 +J\;  m\;\delta m \label{Eq:SuppldeltaE1}\\
&=&2J(1- s_i m)\;.\label{Eq:SuppldeltaE2}
\end{eqnarray}
Note that the in- or decrement of the magnetic moment is $|\delta m|=2 $. According to the Glauber algorithm \cite{glauber1963time} the probability that spin $i$ flips is
\begin{eqnarray}
p_{s_i\to-s_i}&=&\frac{1}{2}\;\frac{\exp(-\beta_{bath}\Delta E/2)}{\cosh(\;\beta_{bath}\Delta E/2\;)}\cr\cr
&=&\frac{1}{1+\exp(\beta_{bath}\Delta E)}\;,\label{Eq:Suppl_transition_flip}
\end{eqnarray}
i.e. vice vera with probability $1-p_{s_i\to-s_i}$ the GFS remains in its state $\sigma$. Translation of this Markov process from the space of microstates $\sigma $ into that of magnetic moments $m$ requires that the random  selection after $\Delta t$ of some arbitrary spin $i$ as a flip candidate is replaced by random selection between a spin down or up polarization. To be consistent with the arbitrary selection of a spin  this random choice is given by the fraction of the respective sorts, i.e. $(N\mp m)/(2N)$ to select a down or up spin, respectively. 
  
For simulation of the stochastic trajectory of the magnetic moment of the GFS, first the polarization was randomly assessed, which determined whether magnetization changed potentially by $\delta m=2$ if down -, or $\delta m=-2$ if up polarization was chosen, respectively. The Glauber's transition probability in Eq.~(\ref{Eq:Suppl_transition_flip})$ $  then determines, whether the spin flip process, i.e. change in magnetization, was realized or not. Figure~(\ref{Fig:Suppl2}) shows  a stochastic trajectory over a long period of time ($10^6 \Delta t $), which visits magnetic moments with equilibrium probability derived from Eq.~(\ref{Eq:peqm}) in the main text (Fig.~(\ref{Fig:Suppl3})) .\\

\subsection{Relaxation of a GFS within a heat bath}

The above considerations are easily transferred to simulate the relaxation of a GFS with a given initial energy/magnetic moment within a heat bath. From some starting magnetic moment stochastic trajectories are followed and sampling at points in time reveals the assigned probability distribution. To evaluate the dynamics of relaxation it is useful to derive the Smoluchowski diffusion equation applied in the main text from the simulation approach.  

\begin{figure}
\includegraphics[width=8cm] {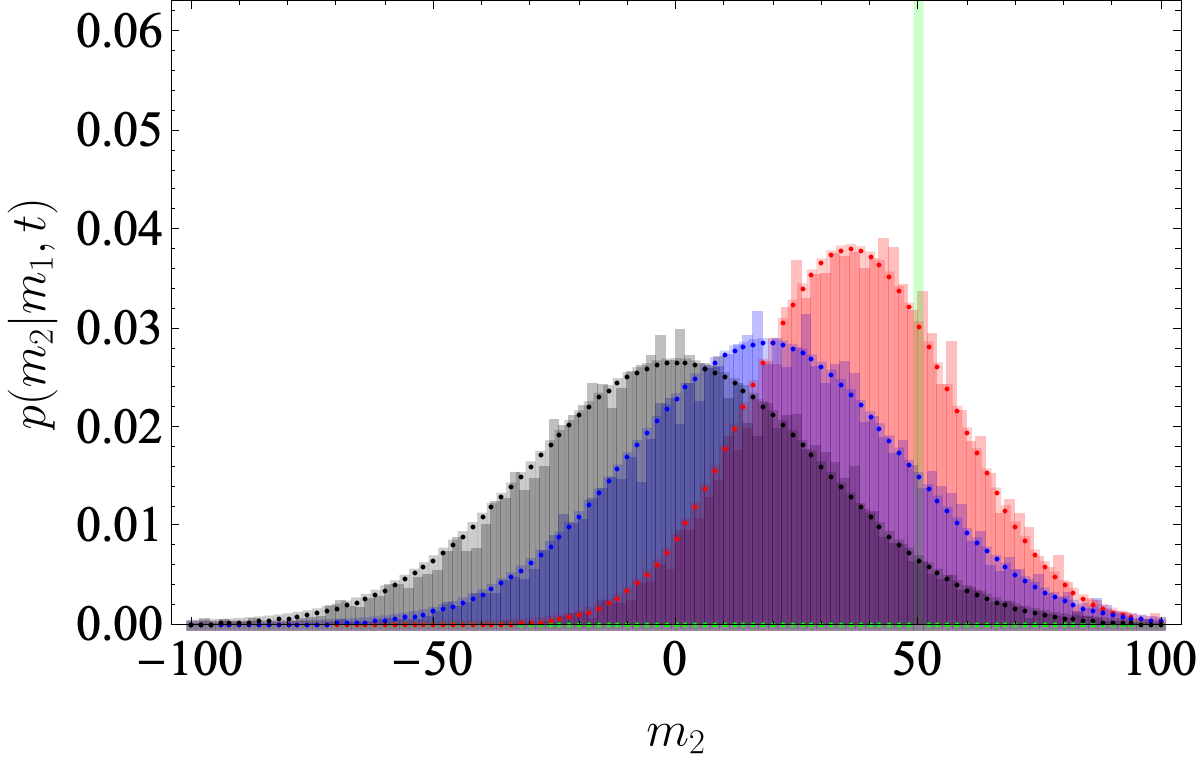}
\caption{Comparison of spin dynamic simulations with analytical results from Eq.~(\ref{Eq:Suppl_Smoluchowski}) for  relaxation of the distribution of the magnetic moment within a heat bath. Consistently with Glauber dynamics the diffusion constant $D$ was chosen unity as well as interaction energy $J$. Parameters of the GFS as spin number $N=1000$, and, hence, its inverse temperature $\kappa=1/(JN)$ and that of the heat bath $\beta_{bath}=1/30^2+\kappa=0.\bar{1} \times 10^{-3}$ are that of Figs.~(\ref{Fig:Suppl2},,\ref{Fig:Suppl3}) and the example in the main text.     8000 trajectories were simulated starting at $m_1=50 $ and followed over 4500 time steps. Four points in time are shown: $ t=0,\;300,\;900, \text{and}\; 4500$, referring to the green, red, blue and grey histograms/curves. With the dynamic constant $\alpha=DJ(\beta_{bath}-\kappa)=1/900 $, which determines speed of relaxation, this refers to $\alpha t =0,\;1/3,\;1,  \text{and}\; 5$ (see main text).} 
\label{Fig:Suppl4}
\end{figure}

The probability that the magnetic moment of the GFS undergoes a stochastic transition to its neighbor value  $m \to m\pm |\delta m| $ after $\Delta t $ is given by Eq.~(\ref{Eq:Supp_transition_m}) and considerations above about selection of spin polarization, 
\begin{equation}
w_{m\to m\pm |\delta m|}=\underbrace{ \frac{N \mp m}{2N}}_{\substack{\text{fraction of} \\ \text{down/up spins  }}}\;\underbrace{\frac{1}{1+\exp\left(\beta_{bath}\Delta E_{m\to m\pm |\delta m|}\right)}}_{\text{Glauber factor}}\;,\label{Eq:Supp_transition_m}
\end{equation}
with $\Delta E_{m\to m\pm |\delta m|}=E_{m\pm|\delta m|}-E_m=J(1/2 \delta m^2\pm  m |\delta m| )$ as the energy difference between final and inital state. Similarly the transition probability from the neighboring moment values to $m $, i.e.  $m\pm |\delta m|\to m $, are obtained. The corresponding hopping rates between nearest neighbors are given by $w/\Delta t $.
The evolution of the probability distribution $p(m,t)$ may be written in form of a master equation with the rates given above, 
\begin{widetext}
\begin{eqnarray}
\frac{\Delta p(m,t)}{\Delta t}=&&\frac{1}{\Delta t}\bigg( w_{m+|\delta m|\to m}\;\; p(m+|\delta m|,t)
+ w_{m-|\delta m|\to m} \;\;p(m-|\delta m|,t)\cr\cr
&-&(w_{m\to m+|\delta m|}+w_{m\to m- |\delta m|})\;\;p(m,t)\bigg).\label{Eq:Suppl_MasterEq}
\end{eqnarray}
\end{widetext}
To approximate the above Equation in the continuum limit, i.e. $|\delta m| \to 0$, $\Delta t\to 0 $ and   replacement of  the discrete probability $p$ by the density  $\rho $, one must consider, that variations of the magnetic moment are partitioned in units of 2. Hence, the appropriate differentials are $\delta m\to 2\;dm$ and $\Delta t\to dt $. Inserting into Eq.~(\ref{Eq:Suppl_MasterEq}) and expanding the terms on the right to the second order $(dm)^2$ makes the Master Equation a Smoluchowski diffusion equation where harmonic forces $F(m)=-\partial_m \left(\tfrac{1}{2}J\;(\beta_{bath}-\kappa)m^2\right) $ determine the drift term, i.e. with $\kappa=-1/(JN) $ as the negative temperature of the GFS, and diffusion coefficient $D $
\begin{widetext}
\begin{equation}
\partial_t\rho(m,t)=\underbrace{\frac{(dm)^2}{dt}}_{=D}\bigg(\partial_m^2\rho(m,t)-\partial_m\bigg(\underbrace{-J\;\left(\beta_{bath}-\kappa \right)m}_{=F(m)} \;\rho(m,t)\bigg)\bigg)\;.\label{Eq:Suppl_Smoluchowski}
\end{equation}   
\end{widetext}
The Greens function, i.e. evolution of the conditional probability density of this Equation, is a shifting Gaussian, starting as a delta distribution ($t=0$) and asymptotically ($t\to \infty$) ending as the Gaussian equilibrium density (see main text).  The discrete probability distribution is regained from density by $p(m)=\int_{m-1}^{m+1}d\tilde{m}\;\rho(\tilde{m}) $.  
Figure~(\ref{Fig:Suppl4}) shows the congruence of relaxation of probability distributions obtained from spin dynamic simulations and the analytical solution (see main text).   

\end{appendix}
%\bibliography{literature}
\end{document}